\documentclass[14pt]{article}
\pdfoutput=1
\usepackage{amsmath,amssymb,amsfonts,amsthm,bm,bbm,cancel,wasysym}
\usepackage{epsfig,graphics,graphicx,epstopdf}
\usepackage{array,booktabs,colortbl,colordvi,multirow}
\usepackage{colordvi,color,xcolor}
\usepackage{hyperref}
\usepackage{rotating}

\usepackage{verbatim}
\usepackage{cite}
\usepackage{subfig}
\usepackage{setspace}
\usepackage{url}
\usepackage[percent]{overpic}
\usepackage{slashed}
\usepackage{authblk}
\usepackage{xspace}
\usepackage{fullpage}
\usepackage{hyperref}

\def\ie{{\it i.e.}}
\def\eg{{\it e.g.}}
\def\etc{{\it etc}}

\def\to{\rightarrow}

\newskip\zatskip \zatskip=0pt plus0pt minus0pt
\def\matth{\mathsurround=0pt}
\def\lsim{\mathrel{\mathpalette\atversim<}}
\def\gsim{\mathrel{\mathpalette\atversim>}}
\def\atversim#1#2{\lower0.7ex\vbox{\baselineskip\zatskip\lineskip\zatskip
  \lineskiplimit 0pt\ialign{$\matth#1\hfil##\hfil$\crcr#2\crcr\sim\crcr}}}

\begin{document}


\begin{flushright}
SLAC-PUB-24003\\
\today
\end{flushright}
\vspace*{5mm}

\renewcommand{\thefootnote}{\fnsymbol{footnote}}
\setcounter{footnote}{1}

\begin{center}

{\Large {\bf Towards UV-Models of Kinetic Mixing and Portal Matter VII: A Light Dark Photon in the $3_c3_L1_A1_B$ Model}}\\

\vspace*{0.75cm}

{\bf Thomas G. Rizzo}~\footnote{rizzo@slac.stanford.edu}

\vspace{0.5cm}

{SLAC National Accelerator Laboratory}\\ 
{2575 Sand Hill Rd., Menlo Park, CA, 94025 USA}

\end{center}
\vspace{.5cm}


\begin{abstract}
\noindent  

The kinetic mixing (KM) portal, by which the Standard Model (SM) photon mixes with a light dark photon arising from a new $U(1)_D$ gauge group, allows for the possibility of viable scenarios of 
sub-GeV thermal dark matter (DM) with appropriately suppressed couplings to the SM. This KM can only occur if particles having both SM and dark quantum numbers, here termed 
portal matter (PM), also exist. The presence of such types of states and the strong suggestion of a need to embed $U(1)_D$ into a non-abelian gauge structure not too far above the TeV scale based 
on the RGE running of the $U(1)_D$ gauge coupling is potentially indicative of an enlarged group linking together the visible and dark sectors. The gauge group 
$G=SU(3)_c\times SU(3)_L\times U(1)_A\times U(1)_B=3_c3_L1_A1_B$  is perhaps the simplest setup wherein the SM and dark interactions are partially unified in a non-abelian fashion that is not 
a simple product group of the form $G=G_{SM}\times G_D$ encountered frequently in earlier work. The present paper describes the implications and phenomenology of this type of setup.

\end{abstract}

\vspace{0.5cm}
\renewcommand{\thefootnote}{\arabic{footnote}}
\setcounter{footnote}{0}
\thispagestyle{empty}
\vfill
\newpage
\setcounter{page}{1}



\section{Introduction and Overview}

As anomalies come and go, the Standard Model (SM) continues to be in exceptional agreement with almost all experimental data.  However, many mysteries still remain unexplained with perhaps 
one of the most obvious being the 
nature of Dark Matter (DM).  As of today, the existence of DM is only known of through its gravitational interactions derived from astrophysical and cosmological observations and whether or not it 
has any further interactions with the SM remains unclear.  If DM indeeds consist of (a set of) new particles (and not, \eg, primordial black holes\cite{Carr:2020xqk}), then the measured value of its relic 
density\cite{Planck:2018vyg} does suggest that some additional, yet unobserved, interactions with at least some subset of the familiar SM particles should be present. Models of particle DM have a 
long history beginning with either/both the QCD axion\cite{Kawasaki:2013ae,Graham:2015ouw,Irastorza:2018dyq} and thermal WIMPS in the few GeV to $\sim 100$ TeV mass 
range\cite{Arcadi:2017kky,Roszkowski:2017nbc,Arcadi:2024ukq} and the broad searches for states such as these continues to push ever further into new parameter space 
regimes\cite{LHC,Aprile:2018dbl,Fermi-LAT:2016uux,Amole:2019fdf,LZ:2022ufs,PandaX:2024qfu,SuperCDMS:2024yiv,Aprile:2024xqi}. However, the so far null results from these searches 
have inspired a host of new potential candidate DM models now known to populate an extremely wide spectrum in both mass and coupling that is found to be dauntingly large 
\cite{Alexander:2016aln,Battaglieri:2017aum,Bertone:2018krk,Cooley:2022ufh,Boveia:2022syt,Schuster:2021mlr,Cirelli:2024ssz}. Attempts to cover even a fraction of this space in the coming years 
through a variety of different novel search techniques will be quite challenging. In a reasonably large fraction of this space, however,  the new interaction(s) between DM and the SM can 
be described by a set of `portals', \ie, effective field theories (EFT) all of which predict the existence of a new class of mediator particles\cite{Lanfranchi:2020crw]}beyond just the DM itself, 
and which may or may not be renormalizable depending upon the specific scenario.

One of the most attractive of these general frameworks is an extension of the thermal WIMP\cite{Steigman:2015hda,Saikawa:2020swg} idea into the few MeV to $\sim 1$ GeV mass range which 
is made possible by introducing the renormalizable Kinetic Mixing/vector portal setup\cite{KM,vectorportal,Gherghetta:2019coi}.  In these models, the DM field carries a 
`dark charge', $Q_D\neq 0$, that is linked to a new gauge interaction - most simply a new (dark) $U(1)_D$ - which has an associated gauge boson, called the dark photon (DP), $V$
\cite{Fabbrichesi:2020wbt,Graham:2021ggy,Barducci:2021egn}. It is assumed that under this new $U(1)_D$ group the SM fields will all have $Q_D=0$ and so the only way for  
the SM and DM fields to interact at low energy scales is via the kinetic mixing (KM) of the DP with SM photon, $A$. Below we will assume that $U(1)_D$ is broken so both the DM and the DP 
obtain their masses (at least partially in the case of DM) via the vev(s) of a, or several, dark Higgs (DH) fields\cite{Li:2024wqj}.  Clearly, in order to construct the vacuum polarization-like diagram(s) 
that are needed to generated this KM, new heavy particles which carry both SM as well as $U(1)_D$ quantum numbers must exist which are either complex scalars and/or vector like fermions (VLF) 
\cite{CarcamoHernandez:2023wzf,CMS:2024bni,Alves:2023ufm,Banerjee:2024zvg,Guedes:2021oqx,Adhikary:2024esf,Benbrik:2024fku,Albergaria:2024pji} so 
as to avoid the well-known unitarity, precision electroweak 
and Higgs boson coupling constraints. We will refer to such exotic particles as Portal Matter (PM) and they have been the subject of much recent attention\cite{Rizzo:2018vlb,Rueter:2019wdf,Kim:2019oyh,Rueter:2020qhf,Wojcik:2020wgm,Rizzo:2021lob,Rizzo:2022qan,Wojcik:2022rtk,Rizzo:2022jti,Rizzo:2022lpm,Wojcik:2022woa,Carvunis:2022yur,Verma:2022nyd,Rizzo:2023qbj,Wojcik:2023ggt,Rizzo:2023kvy,Rizzo:2023djp,Rizzo:2024bhn,Ardu:2024bxg}. 

Given a set of such PM particles, the loop-induced strength of the KM can be expressed via the value of the familiar small dimensionless parameter, $\epsilon$, as 
\begin{equation}
\epsilon =\frac{g_D e}{12\pi^2} \sum_i ~(\eta_i N_{c_i}Q_{em_i}Q_{D_i})~ ln \frac{m^2_i}{\mu^2}\,.
\end{equation}
Here, $g_D$ is identified as the $U(1)_D$ gauge coupling (so that we can likewise define $\alpha_D=g_D^2/4\pi$) and $m_i(Q_{em_i},Q_{D_i}, N_{c_i})$ are the mass (electric charge, dark 
charge, number of colors) of the $i^{th}$ PM field. Also, one has that $\eta_i=1(1/4)$ if the PM is a VLF (complex scalar). In all generality, $\epsilon$ is not a finite quantity {\it but} we might 
imagine that in an at least partially UV-complete theory that group theoretical constraints render the sum 
\begin{equation}
\sum_i ~(\eta_i N_{c_i} Q_{em_i}Q_{D_i})=0\,,
\end{equation}
so that $\epsilon$ will become both finite as well as, in principle, calculable in such a setup. Numerically, for a set of PM fields which have similar masses, such as those we will encounter below, 
we might expect that, roughly speaking, $\epsilon$ lies in the range of $\sim 10^{-(3-4)}$ so that experimental search constraints can be satisfied while also yielding the observed DM abundance for 
sub-GeV thermal DM with similar DP masses.  It is important to note that once one postulates the necessity of PM particles to generated KM, their existence can also induce other loop-level 
interactions between (some of) the SM fields and DM which have a strength similar to that induced by KM. 

As is well-known, there are numerous constraints on these types of setups ranging from direct detection experiments and accelerators searches to cosmology and astrophysics. For example, in 
addition to the thermal DM velocity-weighted annihilation cross section at freeze-out 
requirement\cite{Steigman:2015hda,Saikawa:2020swg}), $\sigma v_{rel}\sim 3\times 10^{-26}~$cm$^3$ sec$^{-1}$, the corresponding cross section must be significantly suppressed at 
later times, \ie, during the CMB\cite{Planck:2018vyg,Slatyer:2015jla,Liu:2016cnk,Leane:2018kjk,Wang:2025tdx} and today\cite{Koechler:2023ual,DelaTorreLuque:2023cef}, implying that such reactions 
must display a significant temperature ($T$) dependence.  This implies that the annihilation through the DP cannot be via a typical $s$-wave process and so the DM cannot be an ordinary 
Dirac fermion. Instead such processes must be, \eg,  ($i$) $p-$wave so that there is a $v^2 \sim T$ suppression at later times, as may be realized in the case of complex scalar or Majorana 
fermionic DM{\footnote {See, however\cite{Belanger:2024bro}}} or may take place $(ii)$ through the co-annihilation mechanism, as can be the case with pseudo-Dirac DM with a sizable mass 
splitting (induced by, \eg, a $Q_D=2$ dark Higgs vev), so that annihilation rate is exponentially Boltzmann-suppressed at later times due to the much lower temperatures\cite{Brahma:2023psr,Balan:2024cmq,Garcia:2024uwf,Mohlabeng:2024itu}. 
Interestingly, especially in this later case, it has been observed that over a significant range of low energy couplings, the RGE running of $\alpha_D$ into the UV indicates that the scale of 
additional new physics, is possibly not very far away due to the eventual loss of perturbativity. This may take place at a scale as low as $\sim 10$ TeV mass or so 
\cite{Davoudiasl:2015hxa,Reilly:2023frg,Rizzo:2022qan,Rizzo:2022lpm}, 
depending, of course, upon the specific details of the low energy field content. The solution to the non-perturbative coupling issue would then be the embedding of the abelian $U(1)_D$ into 
a larger non-abelian, asymptotically free group, $G_D$, which would then reverse the `bad' RGE running {\it before} this high mass scale is reached thus avoiding this problem.  

This observation, combined with the existence of PM - which we can easily imagine might obtain masses as part of the process of $G_D\to U(1)_D$ symmetry breaking - leads to considerations 
of how this new physics might fit together with the SM in a more unified picture, something which has been the subject of much of our recent 
work\cite{Rizzo:2018vlb,Rueter:2019wdf,Rueter:2020qhf,Wojcik:2020wgm,Rizzo:2021lob,Rizzo:2022qan,Rizzo:2022jti,Rizzo:2022lpm,Rizzo:2023qbj,Rizzo:2023kvy,Rizzo:2023djp,Rizzo:2024bhn}, 
with the eventual goal being the construction of an at least partially unified UV-complete setup. Following a bottom-up approach, we have previously examined scenarios of the general product form  
$G_{SM}\times G_D$ where in the simplest setup considered, it has been assumed that $G_D=SU(2)_I\times U(1)_{Y_I}$\cite{Bauer:2022nwt}, into which $U(1)_D$ can be straightforwardly 
embedded in a quite familiar fashion. In this class of setups, although the SM $SU(2)_L$ and the dark $SU(2)_I$ act on orthogonal spaces, at least some of the PM fields lie in $SU(2)_I$ doublet 
representations along with SM fields with which they share common strong and electroweak properties{\footnote {This was motivated by our study in earlier work of $E_6$-type gauge 
models\cite{Hewett:1988xc}}}. In such a setup, additional tree-level interactions will exist between the PM and SM fields due to the exchange of the new heavy non-hermitian gauge boson (NHGB) 
fields in $G_D$ - interactions beyond those arising just from KM. While such scenarios provide us with important insights into more UV-complete setups, they lack any sort of direct (or even partial) 
linkage  between the SM and the dark sector interactions that we'd need to understand better, even if only within the context of a semi-realistic toy model.

In this paper, we will explore a quite different type of semi-unified structure wherein (at least part of) the electroweak interactions of the SM and the $G_D=SU(2)_I\times U(1)_{Y_I}$-type interactions 
linking the SM and PM are described by a {\it single} non-abelian group whose breaking (eventually) will also lead to a light DP. Perhaps the simplest example of a model of this kind wherein the SM 
and these new NHGB-induced interactions arise from a single group that we will consider below is based on the gauge structure $SU(3)_c\times SU(3)_L\times U(1)_A\times U(1)_B=3_c3_L1_A1_B$\cite{Dong:2013wca,Dong:2014wsa,Huong:2016ybt,Leite:2020bnb,VanLoi:2020xcq,Dias:2022hbu,VanDong:2023xmf,Luong:2024xhp,Dong:2015jxa,NguyenTuan:2020xls}. This group has previously 
been examined within the framework of, \eg,  $\lsim$ TeV scale DM and a corresponding massive DP, as well as in other interesting contexts, and which might, in the set of models to be discussed 
below, be regarded as a (very) descoped version of the Quartification class of setups examined previously in Ref.\cite{Rizzo:2023kvy}. 
Models based on this gauge group are themselves extensions of the somewhat more familiar and quite well-studied $3_c3_L1_X$ scenarios\cite{Singer:1980sw,Valle:1983dk,Pisano:1992bxx,
Frampton:1992wt,Foot:1992rh,Montero:1992jk,Foot:1994ym,Long:1995ctv,Tonasse:1996cx,Long:1996rfd,Nguyen:1998ui,Sanchez:2001ua,Martinez:2001mu,Dong:2005ebq,Benavides:2009cn,Deppisch:2016jzl,Benavides:2021pqx,Oliveira:2022vjo,Suarez:2023ozu,Huitu:2024nap,Coutinho:2013lta,Ponce:2002sg,Salazar:2015gxa,Long:2024gyy}, but which is not large enough to contain both a 
light DP as well 
as the direct link between PM and the usual SM fields that we seek here. Note that in these $3_c3_L1_A1_B$ setups the familiar $U(1)_Y$ and $U(1)_D$ are now both partially contained within the 
$1_A1_B$ group factors and we imagine that at some scale, $\gsim 10$ TeV, one (or more) Higgs vev induce the breaking $3_L1_A1_B\to 2_L1_Y1_D$, generating both the PM and new heavy 
gauge boson masses in the process.

The outline of this paper is as follows:  Following this Introduction, in Section 2,  we present the basic overall group/gauge structure of our $3_c3_L1_A1_B$ setup in generality and review its basic 
components from the perspective of a sub-GeV DP and DM but with multi-TeV scale PM. This includes a discussion of anomaly cancellations and parameter values as well as choices of the 
various fermion representations along with the corresponding possible PM electroweak and dark charge assignments. In Section 3, we discuss the three step symmetry breaking chain 
$3_L1_A1_B \to 2_L1_Y1_D\to 1_D1_{em}\to 1{em}$ for both the hermitian and non-hermitian gauge boson sectors and determine the masses and coupling of the new non-SM states in terms 
of the high scale model parameters and the set of required Higgs vevs.  Some simple implications of this symmetry breaking relating the heavy gauge boson masses to each other and the impact 
of their mixings with those of the SM are discussed in Section 4 along with a restriction on the size of the $U(1)_D$ gauge coupling, $g_D$. In Section 5, the masses of the various PM fields are 
analyzed along with their mixing with those of the SM. These mixings are usually critical in allowing for the PM to decay into SM states along with a DP.  In Section 6, the production and decays of 
the PM fields and the new heavy gauge bosons at the LHC and the FCC-hh are then analyzed. Finally, a discussion and our conclusions can be found in Section 7.


\section{Review, Preliminaries and Generic Model Framework }

As noted above, the well-studied $G=SU(3)_c\times SU(3)_L\times U(1)_A \times U(1)_B=3_c3_L1_A1_B$\cite{Dong:2013wca,Dong:2014wsa,Huong:2016ybt,Leite:2020bnb,VanLoi:2020xcq,Dias:2022hbu,VanDong:2023xmf,Luong:2024xhp,Dong:2015jxa,NguyenTuan:2020xls} general framework of models 
are obviously extensions of the even more well-studied $3_c3_L1_X$ model class\cite{Singer:1980sw,Valle:1983dk,Pisano:1992bxx,Frampton:1992wt,Foot:1992rh,Montero:1992jk,Foot:1994ym,Long:1995ctv,Tonasse:1996cx,Long:1996rfd,Nguyen:1998ui,Sanchez:2001ua,Martinez:2001mu,Dong:2005ebq,Benavides:2009cn,Deppisch:2016jzl,Benavides:2021pqx,Oliveira:2022vjo,Suarez:2023ozu,Huitu:2024nap,Coutinho:2013lta,Ponce:2002sg,Salazar:2015gxa,Long:2024gyy}, from which we can extract much 
important model-building information. The matter content of such models consists of a set of $SU(3)_L$ triplets, ${\bf 3}$, and/or anti-triplets ${\bf 3^*}$ of fermion fields, together with a set of 
singlets, ${\bf 1}$, chosen such they contain all of the familiar SM fields in their appropriate representations of the $SU(2)_L$ isospin subgroup, \ie, the ordinary $SU(2)_L$ SM isodoublets, are directly 
embedded into these ${\bf 3}$ and/or ${\bf 3^*}$'s. As we will see below, the additional fermion 
in each of these (anti-)triplets will be identified with PM and will prove to be vector-like with respect to the SM interactions while {\it not} being generally vector-like with respect to the larger gauge 
group. Clearly, since these fundamental triplets and anti-triplets, unlike in the case of the $SU(2)_L$ doublets, are inequivalent representations, the most basic constraint on any model building 
one might consider is that $3_L^3$-type gauge anomaly cancellation will require that the number (including color degrees of freedom) of ${\bf 3}$ and ${\bf 3^*}$ must be equal. Still, this cancellation 
can take place in numerous ways but these can be divided into two main categories depending upon whether or not they cancel within each generation or only when we sum over the contributions of 
all three generations. 
In the former case, the leptonic sector of each generation would need to be significantly augmented as,  \eg, each color triplet quark ${\bf 3}$ would need to have its contribution to this anomaly 
cancelled by 3 sets of color singlet lepton ${\bf 3^*}$'s yielding a total of 18 (chiral) triplets plus anti-triplets for 3 generations. In general in these cases, some of the right-handed (RH), color singlet 
leptons will also end up in ${\bf 3}$'s or ${\bf 3^*}$'s. Furthermore, the gauge interactions in such setups could then be trivially arranged so that there are no obvious tree-level flavor changing neutral 
currents (FCNC). As has been recently stressed\cite{Huitu:2024nap}, in the later case, when one also demands that asymptotic freedom above the PM scale be maintained 
then only 3 generations are allowed - a prediction of such setups. Also in such models, all of the corresponding RH-partners of these fermions will lie in $3_L$ singlets. 
However, though these types of setups would have fewer degrees of freedom (only 12 chiral triplets/anti-triplets instead of 18), 
the price for this greater simplicity would be that FCNC at tree-level would naturally be induced in the quark sector through the exchange of a new heavy neutral gauge boson that arises 
from $G_D$ breaking. As has been much discussed in the literature, this happens because, while we can choose all three lepton generations to lie in, \eg, identical ${\bf 3}$'s,  then it must be that 
two of the quark generations would need to be be in ${\bf 3^*}$'s 
while the remaining one would instead be in a ${\bf 3}$. FCNC would then occur for a heavy gauge boson whose couplings would be sensitive to the this difference among the quark generations 
and this would clearly lead to a lower bound on the mass of such new particles. Representative examples of both of these classes of models will be encountered in our discussion below.

More generally, the potential experimental bounds on the masses of the new heavy gauge and scalar fields that can transmit these FCNC in such classes of models would depend upon exactly how 
these FCNC arise in a particular setup, \eg, whether they occur in the $u,d$ quark and/or lepton sector(s), which if any of the generations are treated differently, the nature and size of the mixing 
between the states which causes them to mix and whether or not additional discrete symmetries/mixing textures are present that may reduce their influence. Further, since at least some of the PM 
fields must in general mix with their SM analogs (via the $\sim 1$ GeV $Q_D$-violating vevs to be discussed below) to allow for all of the PM to decay, these PM fields together with the SM $Z$ 
can {\it also} transmit FCNC making the situation even more complex. In addition to those referenced earlier, this problem has been studied by numerous authors\cite{FCNC} who obtain a wide range of 
possible constraints, from only a few TeV to well over 100 TeV, under a multitude various assumptions. However, it seems clear that the present experimental constraints will allow for new gauge 
bosons with masses not far above the $\sim 10$ TeV scale without too much effort. A detailed discussion of this complex subject is, however, beyond the scope of the present paper. 

As we will also consider further below, the breaking of $3_L1_A1_B\to 1_{em}$ will be accomplished in 3 steps, differing by roughly a factor of $\sim 100$ in mass scale:  the first occurs 
through one large vev, $w\gsim 10$ TeV, which breaks $3_L1_A1_B\to 2_L1_Y1_D$, then through a set of SM breaking vevs, $v_{1,2}\sim 100 $ GeV, as usual, and finally via the vevs 
$u_{1,2} \lsim 1$ GeV which break $U(1)_D$, generating the DP mass as well as PM decay paths through mixing with the SM fields.  while all these vevs arise simultaneously when the scalar 
potential is minimized, these mass scales are sufficiently widely separated that we can discuss them in sequential steps and, essentially, independently since they are to a very good approximation 
decoupled. Some care is, however, required to the situations where this approximation is not 
strictly valid.  As will be further discussed, these various symmetry breaking steps (as well as the generation of the Dirac fermion masses) can all be accomplished through the introduction of, 
for simplicity, only three Higgs scalar ${\bf 3}$'s (or ${\bf 3^*}$'s) each with at least one non-zero vev.  Alternatively, one could assign the $1_D$-violating, $u_{1,2}$ vevs to different 
${\bf 3}$'s (or ${\bf 3^*}$'s) representations.

The gauge part of the covariant derivative for the $3_L1_A1_B$ subgroup of $G$, with which we will be most concerned below (since $3_c$ remains unbroken), can be written as (suppressing 
Lorentz indices for simplicity)
\begin{equation}
g_LT_{iL}W_{iL}+g_AX_AV_A+g_BX_BV_B\,,  
\end{equation}
where $g_{L,A,B}$ are the $SU(3)_L,U(1)_A$ and $U(1)_B$ gauge couplings, respectively. Furthermore, for later 
convenience, we can also define the two coupling constant ratios (in analogy with $t_w=\tan \theta_w=g'/g_L$ in the SM) 
\begin{equation}
t_X=g_A/g_L,~~~t_G=g_B/g_L\,.  
\end{equation}
Note that below the $3_L$ breaking scale, $g_L$ can be directly identified with the usual SM $SU(2)_L$ coupling. 
Here, the $W_i$ are the 8 gauge fields of $3_L$ with the $T_{iL}$ being the corresponding $3_L$ generators normalized so that for the fundamental, ${\bf 3}$, representation are just given by 
$\lambda_i/2$, the $\lambda_i$ being the familiar $3\times 3$ Gell-Mann matrices; for the anti-fundamental ${\bf 3^*}$ representation one has instead that $T_i=-\lambda_i^*/2$. Recall that 
both the $T_{3,8L}$ generators are diagonal with $T_{3L}$ being identified with the third component of the weak isospin in the SM. Similarly, $X_{A,B}$ and $V_{A,B}$ are the corresponding 
gauge charges and gauge fields for the $1_{A,B}$ abelian groups.

The electric charge, $Q=T_{3L}+Y/2$ in the SM, can clearly be expressed, in general, as some linear combination of the $3_L1_A1_B$ diagonal generators. However, we are free to perform a rotation  
in $1_A1_B$ space before symmetry breaking so that the $1_B$ generator does not contribute to this sum. Also, since $T_{3L}$ is identified with the usual SM generator and we can freely choose the 
normalization of the $X_A$ charges, we may define $Q$ as just the sum
\begin{equation}
Q=T_{3L}+\frac{\sigma}{\sqrt 3}T_{8L}+X_A=T_{3L}+\frac{Y}{2}\,,  
\end{equation}
with the second equality just being the SM relationship and 
with $\sigma$ being, {\it a priori}, an unknown (integer) parameter given this choice of normalization, but, as is well-known\cite{Singer:1980sw,Valle:1983dk,Pisano:1992bxx,
Frampton:1992wt,Foot:1992rh,Montero:1992jk,Foot:1994ym,Long:1995ctv,Tonasse:1996cx,Long:1996rfd,Nguyen:1998ui,Sanchez:2001ua,Martinez:2001mu,Dong:2005ebq,Benavides:2009cn,Deppisch:2016jzl,Benavides:2021pqx,Oliveira:2022vjo,Suarez:2023ozu,Huitu:2024nap,Coutinho:2013lta,Ponce:2002sg,Salazar:2015gxa}, can only take on a rather 
restricted set of values as we will see below. From the expression above it follows that (at the $G$ breaking scale in the case of running couplings) 
\begin{equation}
\frac{1}{g_Y^2}=\frac{\sigma^2}{3g_L^2}+\frac{1}{g_A^2}\,. 
\end{equation}
Similarly, the dark charge $Q_D$, which we will normalize here to take on integer values, can also be expressed as a linear sum of the diagonal generators. However, since both fields in the usual SM 
doublets, \ie, $(\nu, e)_L^T$ (with $e$ standing in for a generic charged lepton) and $(u,d)_L^T$, have $Q_D=0$, then $Q_D$ cannot depend upon the generator $T_{3L}$. Thus, since the 
normalization of the charges $X_B$ is arbitrary, we can express $Q_D$ in general as 
\begin{equation}
Q_D=\frac{-2\tau}{\sqrt 3}T_{8L}+2\lambda X_A+X_B\,,  
\end{equation}
and, with a further rescaling, we can freely take $\tau=1$ while $\lambda$ will, for the moment, still remain arbitrary. We might expect in a more complete UV setup that the value of the parameter 
$\lambda$ (not to be confused with any of the Gell-Mann matrices) might be derivable from the other quantities.  We can easily invert these relationships and express the $X_{A,B}$ in terms of the 
other generators;  we then see that
\begin{eqnarray}
&~~~~&X_A= \frac{Y}{2}-\frac{\sigma}{\sqrt 3}T_{8L}\nonumber\\
&~~~~&X_B=Q_D-2\lambda \frac{Y}{2}+\frac{2}{\sqrt 3}(1+\lambda \sigma)T_{8L}\,,
 \end{eqnarray} 
which we can employ to simplify some of the expressions for the covariant derivative and/or couplings of the various gauge bosons that we will encounter below.

Since, in all generality,  we will be embedding the familiar SM isodoublets, $(\nu, e)_L^T$ and $(u,d)_L^T$, into ${\bf 3}$ and/or ${\bf 3^*}$ representations, it is useful to contemplate the following  
possible generic basic constructs which will likely appear in any given specific model that we will consider below:
\begin{equation}
q_1=\begin{pmatrix}u \\ d \\ X_1\\ \end{pmatrix}_L,~~~ q_2^*=\begin{pmatrix} d\\ u\\X_2\\ \end{pmatrix}_L,~~~ l_1=\begin{pmatrix}\nu \\ e \\ X_3\\ \end{pmatrix}_L,~~~l_2^*=\begin{pmatrix}e \\ \nu \\ X_4\\ \end{pmatrix}_L\,,
\end{equation}
with $q_1,l_1$ being ${\bf 3}$'s, while $q_2^*,l_2^*$ are instead ${\bf 3^*}$'s, and with the detailed nature of the fermions $X_i$ being yet unspecified except that we will want to identify them as 
PM fields 
having $Q_D \neq 0$. Note that $X_{1,2}$ are necessarily color-triplet quarks in the SM sense while $X_{3,4}$ correspond to color singlet leptons from this same perspective.  
It is then instructive to act with both the  $Q$ and $Q_D$ generators on these 4 representations and require that we recover the usual electric charges for the SM fields and that these same 
fields all have $Q_D=0$. Doing this will then tell us, amongst other things, the possible values of $Q$ and $Q_D$ for the remaining unfamiliar $X_i$ fields. The first result of this procedure is that 
we find that uniquely, independently of the values of either $\sigma$ or $\lambda$, that the dark charges of these states are completely fixed:
\begin{equation}
Q_D(X_1)=Q_D(X_2)=-Q_D(X_3)=-Q_D(X_4)=1\,. 
\end{equation}
The {\it electric} charges of these states, however, and as is well-known in the $3_c3_L1_X$ model literature, will remain dependent upon the value of $\sigma$ though they are, of course, 
$\lambda$-independent and are summarized in Table~\ref{qtab}.  Clearly, as is well-known, arbitrary values of $\sigma$ are excluded and, in fact, only a small set of possibilities are permissible to 
exclude bizarrely 
charged states. We observe, \eg, that if we choose $\sigma=\pm 1$ then the $X_i$ will carry values of $Q$ which are the same as the familiar quarks and leptons of the SM. If instead, one chooses 
$\sigma=\pm 3$, then we see that the $X_{1,2}$ will have $Q=5/3$ or $-4/3$ (the choice of which one is identified with $X_1$ or $X_2$ being dependent upon the sign of $\sigma$) while the $X_{3,4}$ 
will have similarly $Q=1$ or $-2$. Now since we are not 
identifying any of these $X_i$ with DM but instead as PM, we need all of them to be unstable while they all still carry SM charges as well as having $Q_D=\pm 1$.  While some of the heavy $3_L$ 
gauge bosons will link these $X_i$ with the ordinary SM fermions occupying the same multiplet they will not {\it all} be able to decay this way. At the very least, \eg, the lightest of the PM states will 
be stable unless we allow it to have $U(1)_D$-violating interactions. In previous discussions, we observed that it was the mixings of the PM fields with the corresponding SM ones having the {\it same} 
color and electroweak quantum numbers, which occurs via the same $Q_D$ violating Higgs vevs responsible for generating the DP mass, that was responsible for this. This mixing allowed decays of 
the general form PM$\to$ SM+$V,h_D$, where $h_D$ is the dark Higgs, to occur and these were found to be the dominant decay modes for PM in the simplest approaches. In the present case, 
this requirement would seem to exclude 
the possibility of the exotic charged states such as $Q=5/3,-4/3$ as one of the lightest of these might then be stable and so this would restrict us further (or give greater weight) to the choices 
$\sigma=\pm 1$. Although we will not always impose this requirement in the analysis that follows below, since some clever model building might avoid this apparent outcome, we should remain 
mindful of it as it will come in at a later stage in our discussion as all of the heavy charged PM fields must be allowed to 
decay down to SM states by some means.

\begin{table}
\caption{Electric Charges of the $X_i$}\label{qtab}
The electric charges of the new states, $X_i$, independently of the value of $\lambda$, that lie in the ${\bf 3}$ and ${\bf 3^*}$ representations $q_1,q_2^*,l_1,l_2^*$ as discussed in the text. 
\begin{center}
\begin{tabular}{ l c }
\hline
 State  &  Q   \vspace{0.1cm}\\
\hline 
\vspace{.01cm}\\
$X_1$ & $\frac{1}{6}(1-3\sigma)$  \vspace{.2cm}\\
$X_2$ & $\frac{1}{6}(1+3\sigma)$  \vspace{.2cm}\\
$X_3$ & $-\frac{1}{2}(1+\sigma)$  \vspace{.2cm}\\
$X_4$ & $-\frac{1}{2}(1-\sigma)$  \vspace{.2cm}\\
\hline 
\end{tabular}
\end{center}
\end{table} 

How will the DM itself fit into this setup since we recall that it must be both light $\sim 1$ GeV as well as a SM singlet with $|Q_D|=1$? In the fermionic case, there are essentially two possibilities: if 
it lies in a ${\bf 3}$ or ${\bf 3^*}$ with $Q=T_{3L}=0$, it must be the lowest multiplet member analogous to, \eg, one of the $X_{2,4}$'s introduced above when $\sigma=\mp 1$. In such setups the 
DM mass is generally set by the large vev $w$ discussed above thus having a TeV-scale mass\cite{VanDong:2023xmf}, a situation which we are not interested in here. A second possibility is that 
the DM is instead a vector-like fermion (to avoid any anomalies) which is a singlet under $3_L1_A$ having $Q_D=X_B$ which becomes a pseudo-Dirac state via a dark Higgs vev as discussed in 
the Introduction above, and as recently analyzed in, \eg, Ref.\cite{Rizzo:2024bhn}.  If the DM is a complex scalar without a vev, it can again be chosen to be a $3_L1_A$ singlet state but still 
carrying a dark charge $Q_D\neq 0$, and can easily satisfy the cross section bounds coming from the CMB mentioned above while still obtaining the required relic density observed by Planck.

We now turn to a discussion of the symmetry breaking and mass generation issues in this setup.

\section{Three Stages of Gauge Symmetry Breaking}

\subsection{Hermitian Gauge Bosons}

The goal of this Section is to demonstrate that a light DP with a mass $\lsim 1$ GeV is possible in this type of setup, having all of the required properties, along with the SM photon and $Z$ and, \eg,  
a new neutral heavy gauge boson, $Z'_M$, which might be accessible at the HL-LHC or at other future colliders.  Schematically, the plan is to do this in three distinct stages: the first is via a vev, 
$w \sim 10$ TeV, breaking $3_L1_A1_B \to 2_L1_Y1_D$. Since there is a mass gap until the SM breaking scale is reached, we could imagine that in the energy regime below this large breaking 
scale we could imagine writing something like an EFT by taking the field content of the original model, now rewriting it in terms of $2_L1_Y1_D$ representations, \ie, decomposing the $3_L$ 
triplets into $2_L$ singlets and doublets, and removing/integrating out the heavy fields. Then when electroweak scale is reached a pair of vevs, $v_{1,2}$, who would now appear in $2_L$ doublets, 
will break the SM at $\sim 100$ GeV as usual, \ie, $2_L1_Y\to 1_{em}$. Since there is now another large mass gap, below this scale we could again imagine constructing an EFT, consisting only of 
singlets, by integrating out the SM $W$ and $Z$, until finally $1_D$ is broken by a last pair of vevs, $u_{1,2}$, which we will require to be at the scale $\lsim 1$ GeV leaving us with only unbroken QED. 
Though all of these vevs are the result of a single minimization of the scalar potential, their large hierarchy allow us to treat their actions sequentially in most cases.

We first turn to the symmetry breaking for the neutral gauge boson sector; here we will be very generic and assume for simplicity that the only Higgs fields that are present are those that give masses 
to the quarks and leptons to be discussed below.  As noted above and in previous Section, the breaking of $3_L1_A1_B \to 1_{em}$ takes place in three distinct, widely separated stages via a set 
of $3_L$ Higgs triplets/anti-triplets with the first occurring at or above the $\sim 10$ TeV mass scale where $3_L1_A1_B \to 2_L1_Y1_D$, \ie, the unbroken SM plus the usual (and at this point 
massless) DP.  The vev for this breaking, $w$, must be experienced 
by one of the components of a complex scalar, which here we will denote as $\chi=(\chi_1,\chi_2,\chi_3)^T=H_X$, having $Q=T_{3L}=Q_D=0$, as these symmetries must remain unbroken until 
lower mass scales are reached. These restrictions then tell us that it is the lowest member of this triplet/anti-triplet that obtains the vev, $<\chi_3>=w/\sqrt 2$, to avoid breaking $2_L$ and, further, these 
considerations also completely fix the corresponding values of $X_{A,B}$ for $\chi$ in terms of the parameters $\lambda,\sigma$. As we will see later below, $w$ will also end up generating the 
masses of all of the PM fermion fields as required as well as breaking $3_L1_A1_B\to 2_L1_Y1_D$
Interestingly, when $\sigma =\pm 1$, one also finds that another one of the components of 
$\chi$, (\ie, $\chi_1$ or $\chi_2$) but now carrying $Q_D=\pm 1$, depending upon whether $\chi$ is a ${\bf 3}$ or ${\bf 3^*}$, is also electrically neutral, $Q=0$, and which also has $T_{3L}=\pm 1/2$.  
In such a case this component may {\it also} obtain a vev, \eg, $<\chi_{1(2)}>=u_1/\sqrt 2$, but we will require it to be $\lsim 1$ GeV as it leads to a breaking of $U(1)_D$ and generates a DP mass at 
this scale.  

An occurrence of more than one vev in a single scalar field cannot/is forbidden to happen, \eg, in the SM since there no more than a single component of a Higgs scalar representation can be 
electrically neutral due to the relationship $Q_{em}=T_{3L}+Y/2$ and the fact that we wish to avoid the possibility of charge-breaking minima. However, in more general models with enlarged 
gauge groups where $Q_{em}$ is the sum of several generators, it is possible that two or more components of a given scalar representation can be electrically neutral and so obtain vevs 
simultaneously. Perhaps the most well-known example where this happen is the Higgs bi-doublet in the Left-Right Symmetric Model\cite{LRM}, wherein both neutral components obtain vevs 
whose ratio is $\sim 50$ in order to explain the ratio of the top and bottom quark masses at the weak scale. Another such example is provided by the bi-triplet Higgs representation appearing in 
Trinification models\cite{Trin} which obtain three distinct vevs, which in some cases are very widely separated in scale.  We again stress that these multiple distinct vevs are {\it generated} 
simultaneous as part of the minimization of the potential but, due to their hierarchal nature as occurs in the present setup, can be treated as if they occurred sequentially as to their effects. For 
now, since $w>>u_1$, we can safely ignore the effects of $u_1\neq 0$ for the moment but we  will return to it in the later discussion below. 

Now consider the piece of the covariant derivative corresponding to the set of four neutral gauge fields $W_{3L,8L},V_{A,B}$ acting upon the vev of $\chi_3$ which, since $T_{3L}(\chi_3)=0$, can 
be simply written as (note that $W_{3L}$ will not enter here as this vev has $T_{3L}=0$) 
\begin{equation}
\mp \frac{g_Lw}{3\sqrt 2}\Big( \sqrt 3 W_{8L}-\sigma t_XV_A+2t_G(1+\lambda\sigma)V_B\Big)\,, 
\end{equation}
with the overall sign depending upon whether $\chi$ is a ${\bf 3}$ or ${\bf 3^*}$. Now let us define the quantities 
\begin{equation}
C=[3+\sigma^2t_X^2+4t_G^2(1+\lambda\sigma)^2]^{1/2},~~~~C'=(3+\sigma^2t_X^2)^{1/2}\,, 
\end{equation}
so that we can re-write the expression above as
\begin{equation}
\mp \frac{g_Lw}{3\sqrt 2}C~\Bigg[\frac{ C'\Big(\frac{\sqrt 3 W_{8L}-\sigma t_XV_A}{C'}\Big)+2t_G(1+\lambda \sigma)V_B}{C}\Bigg]\,.
\end{equation}
Now we can further introduce the mixing angle factors 
\begin{eqnarray}
&~&c_\lambda=\frac{\sqrt 3}{C'},~~~s_\lambda=\frac{-\sigma t_X}{C'},\nonumber\\
&~&c_\phi=\frac{2t_G(1+\lambda \sigma)}{C},~~~s_\phi=\frac{C'}{C}\,,
 \end{eqnarray} 
with $t_\lambda =s_\lambda/c_\lambda$, \etc, so that we can define the single (normalized) eigenstate that obtains a mass from the vev $w$ as
\begin{equation}
Z'_M=s_\phi(c_\lambda W_{8L}+s_\lambda V_A)+c_\phi V_B\,, 
\end{equation}
with a mass-squared value given by 
\begin{equation}
M_{Z'_M}^2=\frac{g_L^2w^2}{9}~C^2\,. 
\end{equation}
More generally then, we can write the necessary orthogonal transformation and it's inverse as 
\begin{equation}
\begin{pmatrix} K \\ L\\ Z'_M\\ \end{pmatrix}=O\begin{pmatrix} W_{8L} \\ V_A\\ V_B\\\end{pmatrix} ~~{\rm or} ~~ \begin{pmatrix} W_{8L} \\ V_A\\ V_B\\ \end{pmatrix} =O^T\begin{pmatrix} K \\ L\\ Z'_M\\ \end{pmatrix}\,,
\end{equation}
with the two massless states, $K$ and $L$, being orthogonal to $Z'_M$ so that, explicitly, one has
\begin{eqnarray}
&~&W_{8L}=-s_\lambda K+c_\lambda(c_\phi L+s_\phi Z'_M)\nonumber\\
&~&V_A=c_\lambda K+s_\lambda(c_\phi L+s_\phi Z'_M)\nonumber\\
&~&V_B=-s_\phi L+c_\phi Z'_M\,.
\end{eqnarray} 
In terms of these newly defined fields (and reintroducing $W_{3L}$ as well), the couplings of the 4 hermitian gauge bosons can be written compactly as 
\begin{equation}
g_L\Big(T_{3L}W_{3L}+X_K K+X_L L+X_M Z'_M\Big)\,,
\end{equation}
with
\begin{eqnarray}
&~&X_K=c_\lambda t_X \frac{Y}{2}=\tilde X_K \frac{Y}{2}\nonumber\\
&~&X_L=\frac{C'}{C}Q_D+X_{LY}~ \frac{Y}{2}\,,
\end{eqnarray} 
where, for later use, we have defined the quantity 
\begin{equation}
X_{LY}=\frac{2t_G(3\lambda-\sigma t_X^2)}{CC'}\,,
\end{equation}
and where we also obtain the result that 
\begin{equation}
X_M= C~\frac{T_{8L}}{\sqrt 3}+\frac{2t_G^2(1+\lambda \sigma)}{C}~Q_D-\Big[\frac{\sigma t_X^2+4\lambda t_G^2(1+\lambda \sigma)}{C}\Big]~\frac{Y}{2}\,,
\end{equation}
where we've employed the SM relationship $Y/2=Q-T_{3L}$.

The next stage of symmetry breaking (for our hermitian gauge fields) is to, essentially, generate the SM $Z$ mass which will require vevs with $T_{3L}=\pm 1/2=-Y/2$ (since these states are 
electrically neutral) but still having $Q_D=0$ so that $1_D$ is not simultaneously broken at the electroweak scale. As will be further emphasized below, in doing this we can ignore the $Z'_M$ to 
an excellent approximation as $Z-Z'_M$ mixing is suppressed by a factor of $\sim 10^4$ but we will return to this issue in a later discussion. In practice, we will need to employ two distinct Higgs 
representations to generate, \eg,  the $u-$ and $d-$type 
SM fermion masses which, again, will be ${\bf 3}$'s or ${\bf 3^*}$'s of $3_L$, since in the models we will examine below the right-handed quarks will all lie in $3_L$ singlet representations. To be 
specific, we then need to introduce the two Higgs fields: $\eta=H_u$, one of whose two upper components obtains a vev $v_1/\sqrt 2$, which will generally also give mass to SM $u-$type quarks, and 
$\rho=H_d$, one of whose upper two 
components obtains a vev $v_2/\sqrt 2$, which will also generally give masses to the SM $d-$type quarks in complete analogy with the Type-II Two Higgs Doublet Model (THDM). So, \eg,  if it is 
$\eta_1$, which is the top component that gets the vev $\sim v_1$, then it is $\rho_2$, the middle component, which gets the vev $\sim v_2$. 

To proceed further, since $w^2>>v_{1,2}^2$, as noted we can to a very good approximation simply decouple $Z'_M$ and limit ourselves to just the basis $W_{3L},K,L$. Of course, some 
care in required since $X_M$ has 
a term proportional to $Y/2$, so that both $v_{1,2}\neq 0$ will also induce  $Z-Z'_M$ mass mixing besides generating the SM $Z$ mass, a subject that we will return to later as noted earlier. Acting 
on these states, the covariant derivative can be written symbolically using the definitions above as 
\begin{equation}
D[<\eta>,<\rho>]=g_L(T_{3L}W_{3L}+X_K K+X_L L)[H_u,H_d] =\frac{g_L}{2\sqrt 2} (W_{3L}-\tilde X_K K -X_{LY} L)[v_1,-v_2]\,,
\end{equation}
where the relative sign reflects that the two vevs have opposite values of $T_{3L}=\pm 1/2$. Now let $v$ stand in for either one of $v_{1,2}$; then the expression, above up to a sign, is just 
\begin{equation}
D=\frac{g_Lv}{2\sqrt 2} N_Z Z,~~~N_Z^2=1+N_I^2,~~~N_I^2=\tilde X_K^2+X_{LY}^2\,,
\end{equation}
from which we see that, including now the contributions of both $v_{1,2}$   
\begin{equation}
M_Z^2=\frac{g_L^2}{4}(v_1^2+v_1^2)N_Z^2\,,
\end{equation}
as we might already have guessed and which looks suspiciously like the SM result that we will reproduce provided that we can identify $N_Z^2=1/c^2$ with $c=c_w$ corresponding to the usual 
weak mixing angle. To see that is indeed the case, we first define
\begin{equation}
\alpha=\frac{\tilde X_K}{N_I}~~~\beta=\frac{X_{LY}}{N_I},~~~\alpha^2+\beta^2=1\,,
\end{equation}
so that
\begin{equation}
\tilde X=\alpha K+\beta L, ~~~ \tilde Y=-\beta K +\alpha L\,,
\end{equation}
and thus we can further define the two orthogonal combinations,
\begin{equation}
Z=cW_{3L}-s\tilde X,~~~\tilde A=sW_{3L}+c\tilde X\,,
\end{equation}
with 
\begin{equation}
c=\frac{1}{N_Z},~~~s=\frac{N_I}{N_Z},~~~c^2+s^2=1\,,
\end{equation}
where we still need to show that $c=c_w$ (so that $s=s_w$). To that end we insert these expressions back into the relevant piece of the covariant derivative at this stage yielding, after some algebra, 
first finding  that
\begin{equation}
t_X^2=\frac{3s^2}{3-(3+\sigma^2) s^2}\,,
\end{equation}
and so we arrive at 
\begin{equation}
T_{3L}W_{3L}+X_K K+X_L L=\Big[\frac{1}{c}(T_{3L}-s^2 Q)-(\beta s)s_\phi t_GQ_D\Big]Z+\Big[sQ+(\beta c)s_\phi t_GQ_D\Big]\tilde A+\alpha s_\phi t_GQ_D \tilde Y\,,
\end{equation}
so that $Z$ indeed couples to the SM fields as expected with the identification $s=s_w$, \etc, but also apparently couples to SM singlet dark sector fields at this stage as well and which we will return 
to later below. Now recall that we have yet to break $U(1)_D$ so that neither $\tilde A$ nor $\tilde Y$ are mass eigenstates. These two states will be seen to mix via this remaining symmetry 
breaking step leaving us with the massless photon and the massive DP. 

We recall from above that two components out of the set of Higgs fields, $\chi,\rho,\eta$, in the case $\sigma=\pm 1$, will have an additional $Q=0$ element but with $Q_D=\pm 1$ depending 
upon whether this Higgs is a ${\bf 3}$ or ${\bf 3^*}$ that may obtain a vev, $\sim u_{1,2}$,  that will break $1_D$; further, one of these vevs, $u_1$, will also have $T_{3L}\neq 0$.  This will {\it not} 
happen in the case of $\sigma = \pm 3$ and we would then need to introduce additional Higgs fields to break $1_D$. To simplify the analysis that follows we will assume for now that indeed 
$\sigma =\pm 1$ except as noted (which we will frequently do). Then, similar to what 
was done for isospin breaking above, since these vevs $u_{1,2} \lsim 1$ GeV, we can decouple the heavy SM $Z$, which is an excellent first approximation since $M_Z>> 0.1-1$ GeV, so that we 
can define the mass eigenstates
\begin{equation}
V=\frac{\alpha \tilde Y +\beta c \tilde A}{N_3},~~~A=\frac{-\beta c \tilde Y+\alpha A}{N_3},~~N_3^2=\alpha^2+\beta^2c^2\,,
\end{equation}
where we identify $(V),A$ with the usual (dark) photon and with the masses of these states that are just given by 
\begin{equation}
M^2_A=0,~~~M^2_V=(g_L s_\phi t_G)^2 (u^2_1+u^2_2)\,.
\end{equation}
These mass eigenstates are then found to couple as 
\begin{equation}
g_L \frac{\alpha}{N_3}sQ A+g_L\Big[N_3 s_\phi t_G Q_D+\frac{sc\beta}{N_3}Q\Big]V\,,
\end{equation}
so that we must identify 
\begin{equation}
e=g_L \frac{\alpha}{N_3}s_w\,,
\end{equation}
with the usual electromagnetic coupling to reproduce the familiar QED/SM expression. This all looks rather good {\it except} that we are still left with $V$ having a coupling to $Q$ and the $Z$ having a 
coupling to $Q_D$ with the common feature that these are both proportional to the parameter $\beta$ but neither of which is phenomenologically acceptable if these are both of order unity. We 
notice that if $\beta=0$ then $\alpha=N_3=1$ and so $e=g_Ls_w$ as usual in the SM and we can then define the $U(1)_D$ gauge coupling to be just
\begin{equation}
g_D=g_L s_\phi t_G\,.
\end{equation}
Interestingly,  these conditions can all be easily achieved simultaneously if we assume the rather simple relationship 
\begin{equation}
\lambda \sigma=t_\lambda^2=\frac{s_\lambda^2}{c_\lambda^2}=\frac{\sigma^2t_X^2}{3}\,,
\end{equation}
with $s_\lambda, c_\lambda$ as defined above, and something that we might expect to appear as a signal for and arise from a much more UV-complete picture such as partial unification in 
Quartification\cite{Rizzo:2023kvy} or even complete unification in $SU(N)$\cite{Rizzo:2022lpm}.  Imposing this condition, up to mass and kinetic mixing corrections which are now all of order 
$\sim M^2_Z/M^2_{Z'_M} \sim M^2_V/M^2_Z \sim \epsilon \sim 10^{-4}$, we would obtain a very SM-like situation - subject to several constraints - but aligned with our hoped for expectations and with 
$M_Z^2=g_L^2(v_1^2+v_2^2)/4c_w^2$ as usual in, \eg, the THDM.

\subsection{Non-Hermitian Gauge Bosons}

We now turn to the parallel analysis for the mass generation for the three non-hermitian gauge bosons that appear in this framework. Fortunately, this is far simpler than in the hermitian gauge 
boson case. 
The symmetry breaking implications for the `off-diagonal', non-hermitian gauge bosons (NHGB) are relatively straightforward to analyze as this breaking is contained entirely within the $3_L$ sector 
of the model. The interactions of such fields can be expressed simply through the off-diagonal matrix acting upon, \eg, ${\bf 3}$ fields, here in particular the Higgs scalars, as (we stress that the 
NHGB field $A$ appearing here is not to be confused with the SM photon above) 
\begin{equation}
g_L T_{iL}W_{iL}|_{off-diagonal}=\frac{g_L}{\sqrt 2}~\begin{pmatrix} 0 & W & A\\W^\dagger & 0 & B\\ A^\dagger & B^\dagger & 0\\ \end{pmatrix}\,,
\end{equation}
where we need to consider the individual contributions of each of the three Higgs fields, $\chi,\rho$ and $\eta$ above to the NHGB mass-squared matrix and then simply combine them. Note that 
while the SM $W$ carries $Q=1$, when 
$\sigma=\pm 1$ we see that one of $A$ or $B$ is electrically neutral while the other also carries a $Q=1$ charge. If we make the alternate choice of $\sigma=\pm 3$, then one of the new NHGB would 
have $Q=2$ instead of being neutral. Further, we see that both $A,B$ will carry a non-zero value of $Q_D$ so that the $W$ might then mix with the $Q=1$ NHGB state via one of the, \eg,  
$1_D$-breaking vevs, $u_{1,2}$. Clearly, in any of these cases we expect that to leading order in the vevs and in the absence of any mixing,  $M^2_{A,B} \simeq g_L^2w^2/4$ while 
$M^2_W \simeq g_L^2(v_1^2+v_2^2)/4$. As an example of these mixing effects, let us be specific and consider the case with $\sigma=-1$ and with the  Higgs in triplets; then using the 
expression above, the mass-squared matrix for the NHGB in the ($W,A,B$)-basis would then become 
\begin{equation}
{\cal M}^2_{NHBG}=\frac{g^2_L}{4}~\begin{pmatrix} v_1^2+v_2^2+u_1^2 & 0 & v_1 u_2+wu_1\\0 & w^2+v_1^2+u_1^2+u_2^2& 0\\ v_1 u_2+wu_1 & 0&w^2+v_2^2+u_2^2\\ \end{pmatrix}\,.
\end{equation}
Here we see that our naive expectations are met and that $W-B$ mixing is dominated (assuming that roughly $u_1 \simeq u_2$) by the product of vevs $wu_1$, since $w>> v_{1,2}$, and that 
the relevant mixing angle would then be $\theta_{WB} \simeq u_1/w \lsim 10^{-4}$, which is phenomenologically negligible for most considerations. The result we have obtained is typical of what one  
would find in the other sample cases. For example, when $\sigma=\pm 3$, the $W$ will still mix with the other $Q=1$ gauge boson via the $1_D$-breaking vevs but in this case leaving the $Q=2$ state 
unmixed.  This mixing is also seen to induce a small downward shift in the expected $W$ mass by roughly the same fractional amount, \ie, $\delta M_W^2/M_W^2 \simeq -u_1/w$.

\section{Some Implications}

\begin{figure}[htbp]
\centerline{\includegraphics[width=5.0in,angle=0]{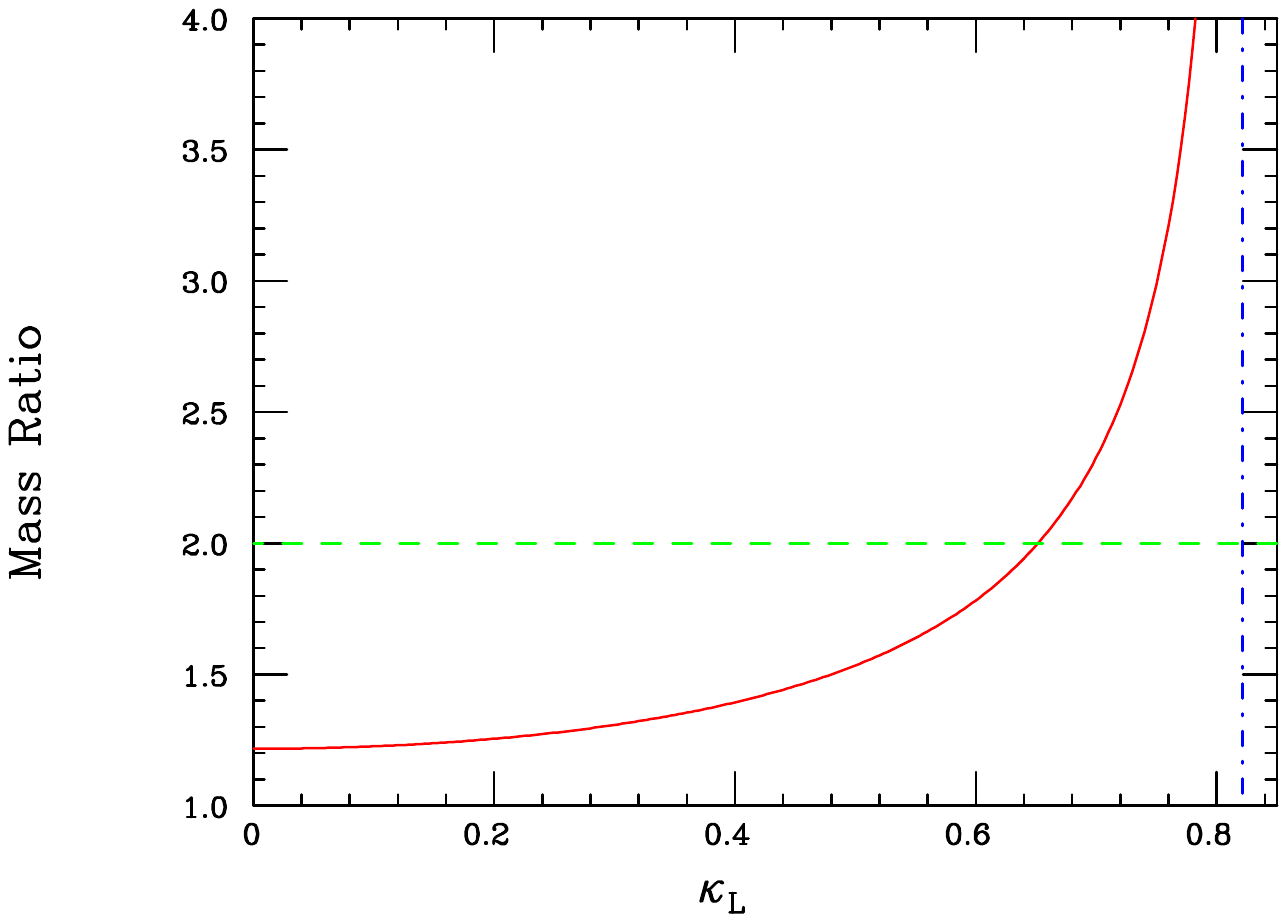}}
\vspace*{-0.8cm}
\centerline{\includegraphics[width=5.0in,angle=0]{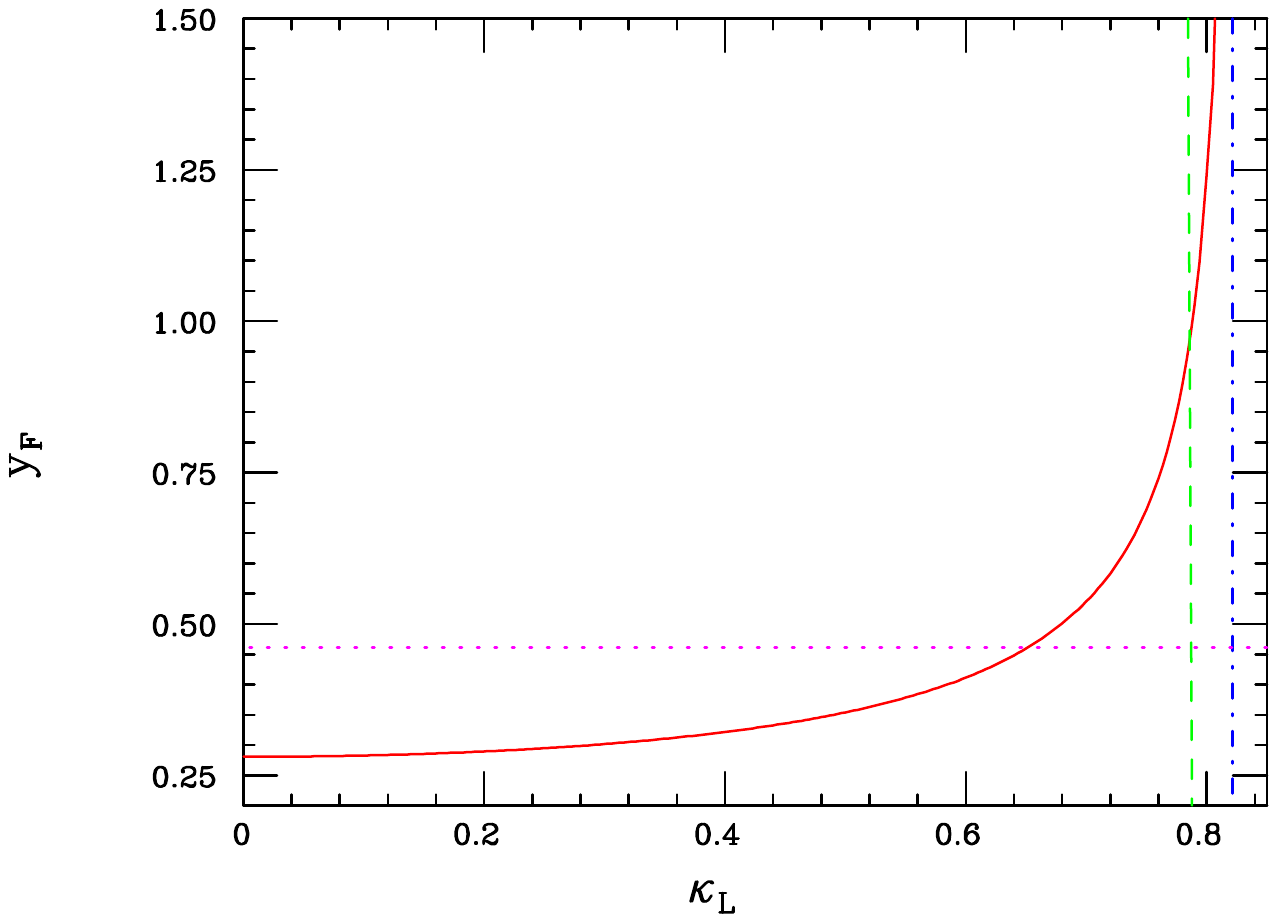}}
\vspace*{-1.3cm}
\caption{(Top) Ratio of the mass of the neutral $Z'_M$ gauge boson to that of either of the the NHGB (solid red curve) as a function of $\kappa_L$ when $|\sigma |=1$ in the $v^2_{1,2}/w^2 \to 0$ 
limit. The vertical dashed-dotted blue line on the righthand side of the Figure show the upper bound on $\kappa_L \simeq 0.821$ in this case as discussed in the text. The horizontal green dashed 
line shows the mass threshold beyond which $Z'_M$ is kinematically allowed to decay to NHGB pairs which occurs when $\kappa_L$ exceeds $\simeq 0.652$.  (Bottom) Maximum value of the 
PM ($F$) generic Yukawa coupling $y_F$ as a function of $\kappa_L$ above which the decay $Z'_M\to \bar FF$ is kinematically forbidden.  The vertical green dashed line on the right hand side 
of the panel corresponds to the value of $\kappa_L\simeq 0.788$ when $y_F=1$. The magenta dotted line indicates the corresponding maximum value of $y_F\simeq 0.462$ beyond which 
NHGB decays to $\bar Ff$ are kinematically forbidden.   As in the upper panel, the blue vertical dash-dotted line shows the upper bound on $\kappa_L$.}
\label{figa}
\end{figure}

If we demand that the relationship $\lambda \sigma=t_\lambda^2$ holds (especially in the cases where $\sigma^2=1$ which we will generally assume henceforth except where noted) then there 
are many simplifications in the expressions above and the results of the previous analysis become much more transparent with a number of interesting implications. A simple example is provided by
Eq.(33) above; defining the abbreviations 
\begin{equation}
\kappa_L=\frac{g_D}{g_L}, ~~~~~~ r=\frac{4(1-x_w)}{[3-(3+\sigma^2)x_w]}\,,
\end{equation}
with $x_w=s_w^2 \simeq 0.2315$ at the weak scale as usual, we find that we can express $t_G$ simply as 
\begin{equation}
t_G^2=\frac{\kappa_L^2}{1-\kappa_L^2r}\geq 0\,,
\end{equation}
implying that $\kappa_L$ is {\it bounded} from above, \ie, 
\begin{equation}
\kappa_L\leq \frac{1}{\sqrt r}\simeq 0.821(0.269)\,,
\end{equation}
for $|\sigma |=1(3)$ or $g_D\lsim 0.535(0.176)$, again employing suggestive weak scale input values. If $g_D$ runs to smaller values as the $\sim 1$ GeV mass scale is approached from above, 
as we might perhaps expect, this can have implications for low energy dark sector searches. Another simple example is provided by the mass ratio of the new heavy hermitian gauge boson, $Z'_M$, 
to those of the new NHGB encountered in the previous Section (in the limit where the sub-leading $v^2_{1,2}$ contributions can be neglected) is just 
\begin{equation}
\frac{M^2_{Z'_M}}{M^2_{NHGB}}=\frac{r}{1-\kappa_L^2 r}\,,
\end{equation}
which is always greater than unity and has important phenomenological implications as it determines whether of not NHGB pairs can be produced resonantly via the $Z'_M$, \ie, when the relation 
$M_{Z'_M}>2M_{NHGB}$ holds. This requirement is shown in the upper panel of Fig.~\ref{figa} in the case when $|\sigma|=1$ and  we observe that $\kappa_L\gsim 0.652$ must be satisfied for such 
processes to occur. Interestingly, when $|\sigma|=3$, this mass ratio requirement is {\it  always} satisfied for all physical $\kappa_L$ values as $r$ in this case is so much larger than when 
$|\sigma|=1$.

Lastly, we consider the case of $Z-Z'_M$ mixing as alluded to previously; this quantity partly arises from the terms in $X_M$,  as introduced above, which are proportional to both $T_{8L}$ and also to 
$Y/2=-T_{3L}$, with the 
last equality holding when acting on the $Q_D=Q=0$ components of the Higgs fields that obtain the electroweak scale vevs, $v_{1,2}$. Note that these vevs have $T_{8L}/\sqrt 3=1/6, T_{3L}=\pm 1/2$, 
respectively. Employing the same notation as above, for simplicity we first introduce the abbreviations
\begin{equation}
a=\frac{3}{2} \Big(\frac{r}{1-r\kappa_L^2}\Big)^{1/2},~~~b=-\frac{1}{4a\sigma}~\Big(r-\frac{4}{3}\Big)~\Big[\frac{1+8r\kappa_L^2}{1-r\kappa_L^2}\Big]\,.
\end{equation}
Note that for $\sigma=\pm 1$, $r\simeq 1.482$ so that $b$ is generally small in comparison to $a$. 
Then, to leading order in the ratio of squared vevs, $v_{1,2}^2/w^2$, the $Z-Z'_M$ mass-squared submatrix is given by
\begin{equation}
{\cal {M}}^2_{Z-Z'_M}\simeq \begin{pmatrix} M^2_{Z'_M} & M_{int}^2 \\ M_{int}^2& M_Z^2 \\  \end{pmatrix}\,,
\end{equation}
where both $M_Z^2$ and $M_{Z'_M}^2$ are given above and $M_{int}^2$ is given in terms of the parameters $a,b$ by 
\begin{equation}
M_{int}^2=\frac{g_L^2}{2c_w}\Big[\frac{b}{2}(v_1^2+v_2^2)+\frac{a}{6}(v_2^2-v_1^2)\Big]\,,
\end{equation}
which leads to an explicit expression for the $Z-Z'_M$ mixing angle to leading order in the vev ratios given by  
\begin{equation}
\theta_{mix} \simeq \frac{M_{int}^2}{M_{Z'_M}^2} =\frac{3}{4ac_w}\Big[\frac{3b}{a} \Big(\frac{v_1^2+v_1^2}{w^2}\Big) +\frac{v_2^2-v_1^2}{w^2}\Big]\sim 10^{-4}\,,
\end{equation}
with the magnitude being as expected. This induces, at the same level of approximation, a small downward fractional shift in the SM $Z$ mass explicitly given by
\begin{equation}
\frac{\delta M_Z^2}{M_Z^2}\simeq -\frac{1}{36w^2}\Big[\Big(1-\frac{3b}{a}\Big)^2v_1^2+\Big(1+\frac{3b}{a}\Big)^2v_2^2\Big]\sim 10^{-4}\,,
\end{equation}
which we see is again of the same small magnitude as found in the case of $W-$NHGB mixing.

\section{Fermion PM Masses}

Given the setup above we can now more explicitly discuss the Dirac masses of the PM fields, all of which lie at the scale of the vev $\sim w$. In the case where the anomalies cancel between the 
fermions amongst the 3 generations, the PM will consist of some set of the $X_i$ introduced above as left-handed ${\bf 3}$'s or ${\bf 3^*}$'s and right-handed singlets. When the anomalies instead 
cancel within each 
of the generations, then the quark (\ie, color triplet) sector will still appear as a set of the $X_{1,2}$ as just described but then the lepton sector is necessarily more complex as was discussed above.  
This being the case, we will begin with the `simpler' quark sector where, ignoring flavor issues, there are two subcases, denoted as $Q=q_1,q_2^*$ above, depending upon whether the $u$ or $d$ 
quark, respectively,  has a PM partner field with which it mixes via one or more of the $Q_D$-violating vevs. For example, taking $\sigma=-1$, if we make the choice the $Q=q_2^*$, we see that 
the quark mass terms can be written as 
\begin{equation}
{\cal L}_{Yukawa}= y_d \bar Qd_R\eta^*+y_u\bar Q d_R \rho^*+y_D\bar QD_R\chi^*+{\rm h.c.}\,,
\end{equation}
which, once all of the vevs are turned on, yields the $2\times 2$ mass matrix
\begin{equation}
{\cal M}_d=\frac{1}{\sqrt 2}\begin{pmatrix} \bar d_L,&\bar D_L \\ \end{pmatrix}  \begin{pmatrix} y_dv_1 & y_D u_2 \\ y_du_1 & y_Dw \\  \end{pmatrix} \begin{pmatrix} d_R \\ D_R \\ \end{pmatrix}\,,
\end{equation}
while $m_u=y_uv_2/\sqrt 2$. ${\cal M}_d$ is, as usual, diagonalized by a bi-unitarity transformation, $M_{diag}= U_L {\cal M}_dU_R^\dagger$ which, if we neglect any phases for simplicity, can 
be parameterized by employing two small mixing angles $\theta_{(L,R)}^d \simeq (u_2, y_du_1/y_D)/w \lsim 10^{-4}$, since we might expect that $y_d/y_D<1$, to leading order in the vev ratios. 
This is roughly the same magnitude as was found for both of the 
$Z-Z'_M$ and $W-$NHGB mixing factors obtained above. Note that if we had instead chosen $Q=q_1$, then 
essentially the roles of $d$ and $u$ would be interchanged with the replacement $D\to U$. Similarly, when the left-handed leptonic multiplet is either of $l_1,l_2^*$, as is the case when the 
anomalies cancel amongst the three generations, an essentially identical set of results are found to hold. As alluded to previously, these rather small mixings between the PM and the corresponding SM 
particles with which it shares a multiplet, are of great phenomenological relevance as they generate the PM dominant decays, such as $D\to dV,h_D$, with partial rates proportional to the combinations  
$\sim (\theta_{L,R}m_D/m_V)^2$ which we see is $\sim O(1)$ as was shown in very early work on PM\cite{Rizzo:2018vlb}. Clearly, this important mixing cannot occur when $X_i$ has an exotic 
electric charge as will be the case when the corresponding choice $|\sigma|=3$ is made.

Of course, in the leptonic sector, the situation is a bit more complex when the gauge anomalies cancel generation by generation due to the required augmentation mentioned above, \ie, three leptonic 
${\bf 3}$'s are needed to cancel the contribution to the $3_L^3$ anomaly of a single quark ${\bf 3^*}$ for each generation and vice-versa. In models such as this, fields with SM charges but with 
$Q_D=\pm 2$ can occur. The simplest examples of this, when $\sigma=1$, are the $S_{9,10}$ 
sets of lepton representations as given in \cite{Benavides:2021pqx} for a single generation that are quite representative of this possibility and so we will examine them briefly here. In the case of $S_9$, 
all of the leptonic fields lie in three ${\bf 3^*}$ representations, $R_{1-3}$, {\it including} the conjugates of the left-handed fields, \eg, $e^c_L$, \etc, which can be written in the form
\begin{equation}
S_9:~R_1=\begin{pmatrix}e^- \\ \nu \\ N_1\\ \end{pmatrix}_L,~~~ R_2=\begin{pmatrix} E\\ N_2\\N_3\\ \end{pmatrix}_L,~~~ R_3=\begin{pmatrix}N_2^c \\ E^c \\ e^c\\ \end{pmatrix}_L\,,
\end{equation}
where we see that the leptonic PM fields consist of a $Q=Q_D=-1$ Dirac state, $E=E^-$, as well as the four neutral fermions, $N_{1-3},N_2^c$, with various transformation properties, \eg, 
$Q_D(N_1)=Q_D(N_2)=-1$ while $Q_D(N_3)=-2$. Given our previous discussion, such a state as $N_3$ would necessarily decay via a NHGB exchange if it doesn't mix with any of the others.  Also 
note that only a conjugate field exists for $N_2$ to form a Dirac mass term.  In such a case, as 
there are no right handed singlets present, the relevant fermion mass terms must be generated via the asymmetric triple product of pairs of the $R$'s, with the set of (conjugated) Higgs 
fields, $H_{A,B,C}^*$ with $H$'s being generally similar to one of the members of the set $(\chi, \eta,\rho)$ introduced in Section 2 above. This results in a set of general interactions of the form
\begin{equation}
{\cal L}_{S_9}=\epsilon_{ijk}\Big(\kappa_AR^i_2R^j_3H_A^{k*}+\kappa_BR^i_1R^j_3H_B^{k*}+\kappa_CR^i_1R^j_2H_C^{k*}\Big)+{\rm h.c.}\,.
\end{equation}
where the $\kappa$'s are a set of Yukawa couplings and with the $(i,j,k)$'s labeling the various components of the relevant representations. Here we observe, for example, that $H_A \sim \chi$ and 
$H_B \sim \eta$ above while we'll see that $H_C$ is necessarily somewhat different.

For the charged leptons, after the Higgs fields obtain their vevs, this set of couplings results in the $2\times 2$ mass matrix
\begin{equation}
{\cal M}_e=\frac{1}{\sqrt 2}\begin{pmatrix} e,&E \\ \end{pmatrix}_L  \begin{pmatrix} -\kappa_Bv_1 & \kappa_Bu_2 \\ -\kappa_Au_1 & \kappa_Aw \\  \end{pmatrix} \begin{pmatrix} e^c \\ E^c\\ \end{pmatrix}_L\,,
\end{equation}
which is quite similar to the matrix ${\cal M}_d$ discussed previously above. We also obtain the following $4\times 4$ Dirac mass matrix for the neutral fields given by
\begin{equation}
{\cal M}_\nu=\frac{1}{\sqrt 2}\begin{pmatrix} \nu,&N_1,&N_2,&N_3 \\ \end{pmatrix}_L\begin{pmatrix} 0 & 0 & -\kappa_Bu_2 &0\\ 0 & 0 & \kappa_Bv_1 & 0 \\0 & 0 & -\kappa_Aw & 0 \\0 & 0 & \kappa_Au_1 & 0\\  \end{pmatrix} \begin{pmatrix} \nu^c \\ N_1^c\\ N_2^c\\N_3^c\\\end{pmatrix}_L\,,
\end{equation}
where we see that only one combination, essentially $N_2N_2^c$ to leading order in the vevs, gets a Dirac mass at this stage as the fields $\nu^c, N_{1,3}^c$ do not appear among the $S_9$ set.  
Two Majorana mass terms are also found to be potentially generated by the last term in these couplings:
\begin{equation}
\frac{\kappa_C\tilde u}{\sqrt 2}\big (\nu N_3-N_1N_2\big)+{\rm h.c.}\,.
\end{equation}
but this requires that the top component of $H_C^*$ with $Q=0$ obtain a small vev, $\tilde u \lsim 1$ GeV, while this same field simultaneously must have $Q_D=2$. This is perhaps not so strange as a 
$Q_D=2$ Higgs vev is also needed to generate pseudo-Dirac dark matter to satisfy the CMB constraints as mentioned in the Introduction. While such a field will contribute to the DP mass, if it 
exists it will not significantly influence the discussion of the gauge boson masses presented in the previous sections. Clearly the neutral lepton sector of this type of setup deserves some further study. 

In the case of $S_{10}$, the leptons of a single generation (and their conjugates) now lie in three different ${\bf 3}$ representations, $R_{1-3}$, as well as three additional $3_L$ singlets which we 
can in terms of left-handed fields as write as:
\begin{equation}
S_{10}:~~R_1=\begin{pmatrix}\nu \\ e \\ E_1\\ \end{pmatrix}_L,~~ R_2=\begin{pmatrix} E_2^c\\ N_1^c\\N_2^c\\ \end{pmatrix}_L,~~~R_3=\begin{pmatrix}N_1 \\ E_2 \\ E_3\\ \end{pmatrix}_L,~~
e^c_L,~E_{1L}^c,~E_{3L}^c\,,
\end{equation}
where we now see that the PM states consist of the three Dirac charged states, $E_{1-3}^-$, as well as two neutral states, $N_{1,2}$, all with various transformation properties. In principle, due to the 
parameter freedom in this model, unusual $Q_D$ quantum numbers can arise in this case. With the SM fields having $Q_D=0$ as usual, we see that the general assignments 
$Q_D(E_1)=1$, $Q_D(E_2,N_1)=q$, $Q_D(N_2^c)=1-q$, and $Q_D(E_3)=1+q$ are possible for arbitrary values of $q$with the choice $q=1$ being the simplest one that we will consider below. 
In such a case we see that we can make the important identification $N_2^c=\nu^c$ so that the SM neutrinos can obtain an ordinary Dirac mass. Here 
the SM and PM masses are not only the result of the triple product of fields, as was the case with $S_9$ above, but are also due to the more familiar couplings of the quark-like 
${\bf {313^*}}$ product variety already encountered above. A minimal set of such couplings (without, \eg,  introducing any additional $|Q_D|=2$ Higgs fields) is given by 
\begin{equation}
{\cal L}_{S_{10}}=\epsilon_{ijk}\Big(y_\nu R_1^i R_2^j H(\nu)^k+y_DR_2^i R_3^j H(N_1,E_2)^k\Big)+ y_eR_1e^cH(e)+y_{E_1}R_1E_1^cH(E_1)+y_{E_3}R_3E_3^cH(E_3)+{\rm h.c.}\,,
\end{equation}
where, as before $i,j,k$ label the multiplet member as before  and the $H$'s are potentially different Higgs representations denoted by the fields to which they give mass. Some algebra tells us, in the 
notation above, that in terms of transformation properties and vevs, $H(E_1)=H(E_3)\sim \chi^*$, $H(e)\sim \eta^*$, $H(N_1,E_2)\sim \chi$, and $H(\nu)\sim \eta$. From this interaction, after the 
all the Higgs fields obtain their vevs, the following $4\times 4$ mass matrix is generated for the $|Q|=1$ fermions:
\begin{equation}
{\cal M}_e=\frac{1}{\sqrt 2}\begin{pmatrix} e,&E_1,&E_2,&E_3 \\ \end{pmatrix}_L\begin{pmatrix} y_ev_1 & y_{E_1}u_2 & -y_\nu u_1 &0\\ y_e u_1 & y_{E_1}w & y_\nu v_1 & 0 \\0 & 0 & y_Dw & y_{E_3}u_2 \\0 & 0 & -y_Du_2 & y_{E_3}w\\  \end{pmatrix} \begin{pmatrix} e^c \\ E_1^c\\ E_2^c\\E_3^c\\\end{pmatrix}_L\,.
\end{equation}
Here we see that the PM fields obtain their usual large masses from the $w$ vev and that both $e-E_1$ and $E_2-E_3$ will mix via the $U(1)_D$-violating vevs $u_{1,2}$ while $E_1-E_2$ mixing is 
generated by the electroweak scale vev, $v_1$. These mixings are observed to be sufficient to allow the $E_i$ to all eventually decay down to the electron plus DP final state as required.  
Similarly, the corresponding $2\times 2$ Dirac mass matrix for the neutral fields, after now making the identification $\nu^c=N_2^c$,  is then given by
\begin{equation}
{\cal M}_\nu=\frac{1}{\sqrt 2}\begin{pmatrix} \nu,&N_1\\ \end{pmatrix}_L\begin{pmatrix} -y_\nu v_1 & -y_\nu u_1 \\ y_Du_2 & -y_Dw  \\  \end{pmatrix} \begin{pmatrix} \nu^c \\ N_1^c\\ \end{pmatrix}_L\,,
\end{equation}
while we see that no Majorana mass terms are generated from just these terms alone without further extending ${\cal L}_{S_{10}}$, which is certainly possible. This is similar to that obtained for $e-E$ 
mixing in the case of the $S_9$ setup previously discussed as well as what was obtained for the quark sector except that $y_\nu$ must be highly suppressed to explain the SM neutrino 
masses while we might expect $y_D\sim O(1)$. This implies that $\theta_L^\nu \simeq (y_\nu /y_D) u_1/w <<\theta_R^\nu \simeq u_2/w \sim 10^{-4}$, but which still allows for the decay $N_1\to \nu V$ 
to occur. 

One of the obvious results of the discussion above is that we find, as expected, that the masses of all of the PM fermions, $F$,  are given generically by a relation of the 
form $m_F\simeq y_Fw/\sqrt 2$, 
to leading order in the vevs, where we might expect that the various $y_F$'s to be $O(1)$. An important question to ask, especially when we address the phenomenological collider signatures below, is 
whether or not the new heavy gauge bosons, $Z'_M$ and the NHGB above, can decay into such states, \ie, $Z'_M\to \bar FF$, NHGB$\to \bar Ff {(\rm or} \bar f F)$ as these would then be 
potentially resonantly enhanced cross sections, particularly in the $Z'_M$ case. The case for the NHGB is simple as, to leading order in the vevs,  $M_{NHGB}\simeq g_Lw/2$ so that we see that 
this decay channel is open provided that $y_F< \sqrt 2(\sqrt 2 G_F M_W^2)^{1/2}=\sqrt 2x\simeq 0.462$, a bound which is shown in the lower panel of Fig.~\ref{figa}. In the case of $Z'_M\to \bar FF$, 
the corresponding bound on $y_F$ will clearly be $\kappa_L$-dependent and is given instead by 

\begin{equation}
y_F< \frac{x}{\sqrt 2}~\Big[\frac{r}{1-r\kappa_L^2}\Big]^{1/2}\\,
\end{equation}
so that, \eg, if $y_F=1$ (similar to that of the top quark on the SM) and $\sigma=\pm 1$, then $\kappa_L>0.788$ is needed for this decay to be kinematically allowed, a value not much smaller that 
the upper bound of $\simeq 0.821$ previously obtained on $\kappa_L$ above. This constraint on $y_F$, as a function of $\kappa_L$, is also shown in the lower panel of Fig.~\ref{figa}. In the case 
where $\sigma =\pm 3$, $r$ is significantly larger by almost an order of magnitude such than even potentially perturbatively troublesome values of $y_F$ would be still allowed as long as we stayed  
away from the corresponding upper bound on the value of $\kappa_L$ in this case, \ie, $\simeq 0.269$. 

\begin{figure}[htbp]
\centerline{\includegraphics[width=5.0in,angle=0]{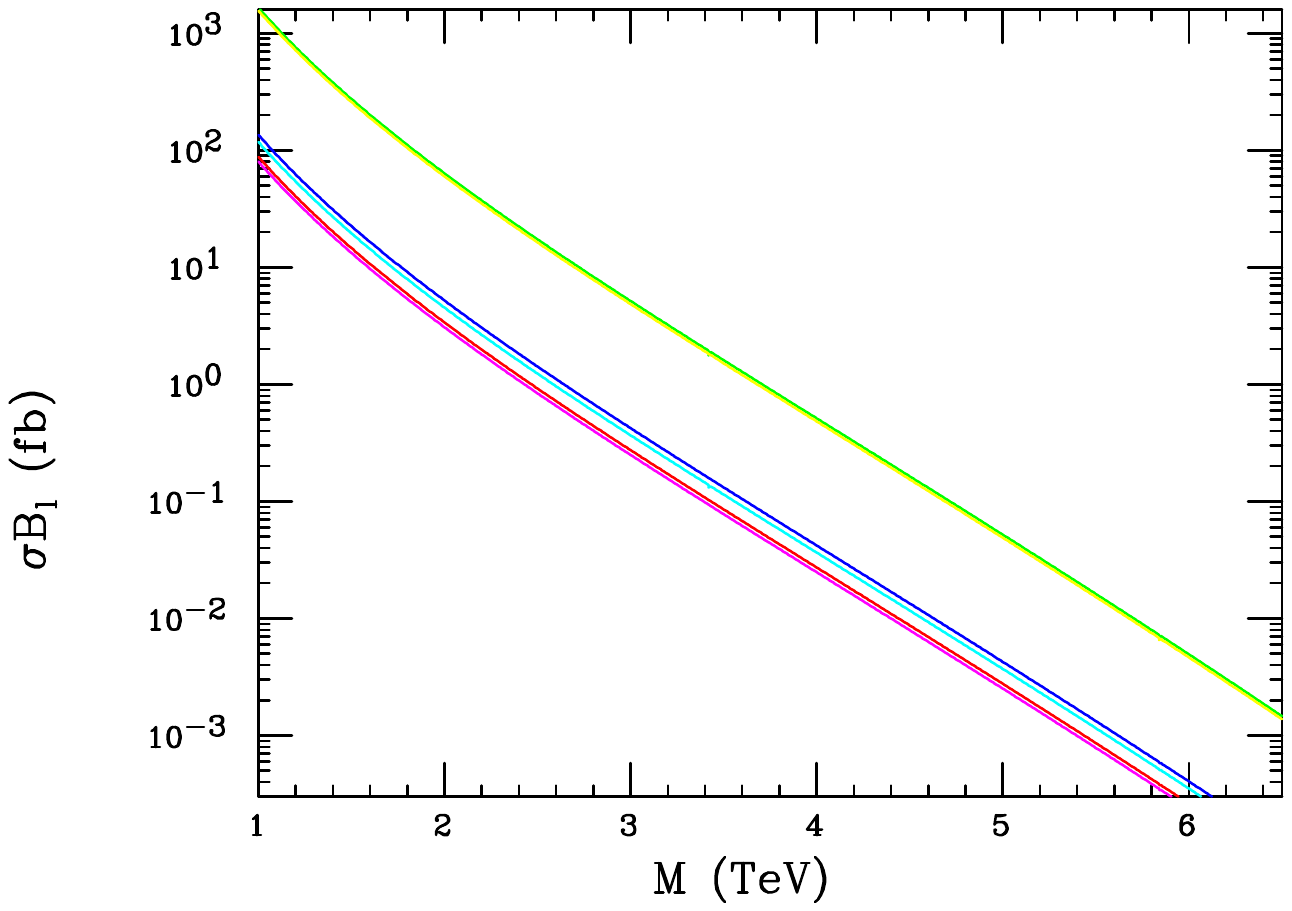}}
\vspace*{-0.8cm}
\centerline{\includegraphics[width=5.0in,angle=0]{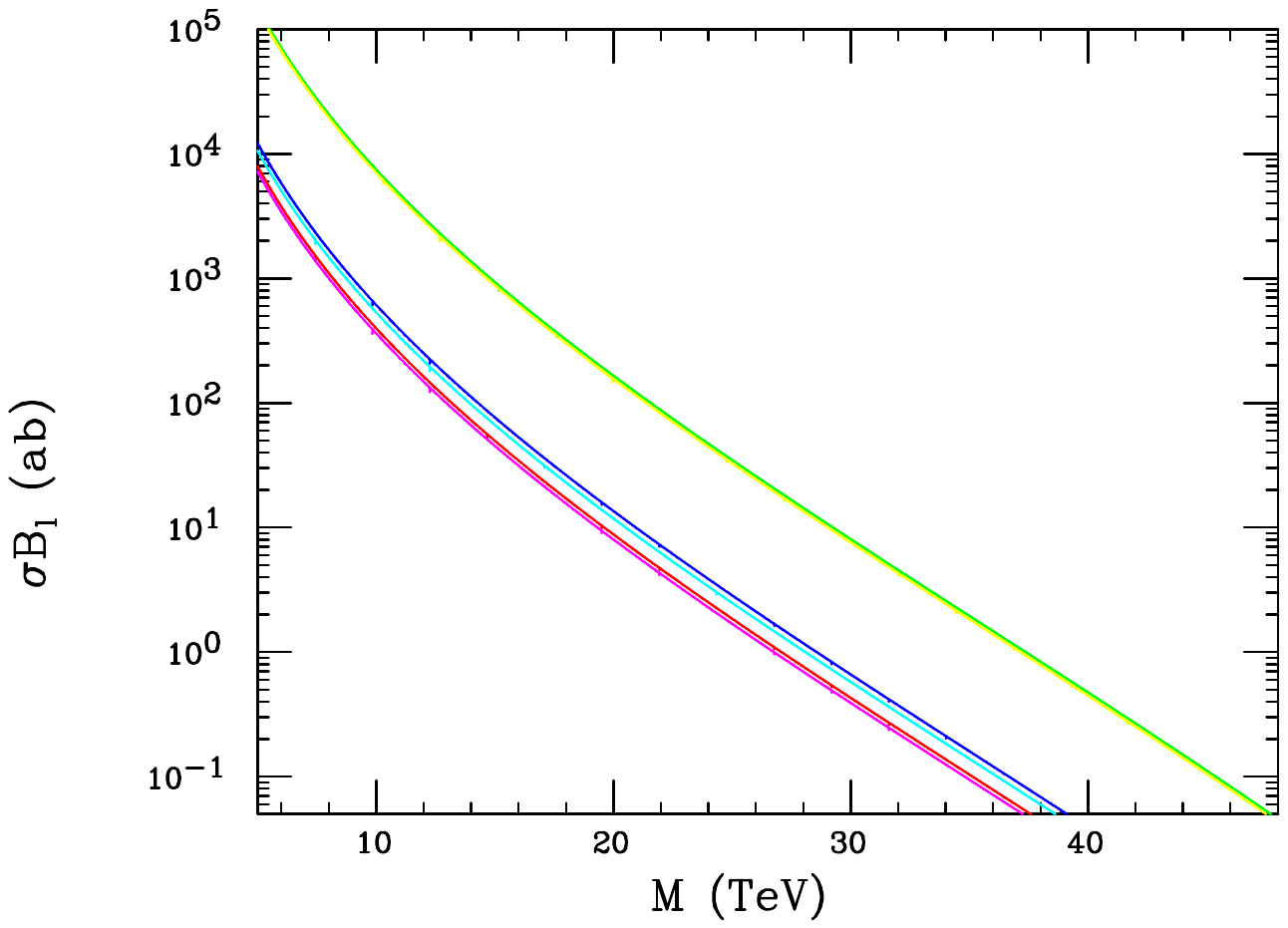}}
\vspace*{-1.3cm}
\caption{Production cross section times leptonic branching fraction for the new heavy gauge boson, $Z'_M$, as a function of its mass employing the Narrow Width Approximation as discussed in the 
text for (Top) the 13 TeV LHC and (Bottom) the 100 TeV FCC-hh. The red (blue, green) curves correspond to $\kappa_L=0.1, \sqrt x_w (\ie, g_D=e),$ and $0.8$, respectively, with the choice of 
$\sigma=1$. The magenta (cyan, yellow) curves are the corresponding results assuming instead that $\sigma=-1$, and we observe very little difference. In all cases only $Z'_M$ decays to SM fermions 
have been assumed to be kinematically accessible.}
\label{figb}
\end{figure}

\section{Collider Signatures From New Heavy States}

Having the results of the analyses above in hand, we can now turn to a discussion of some of the collider implications of this setup resulting from the production of the new heavy gauge bosons as 
well as the PM fields; here we will many times limit ourselves to the cases where $\sigma=\pm 1$ for definiteness but occasionally we will note the differences were we to instead choose $|\sigma|=3$.  
Of course, in addition to new physics at colliders, the existence of these new states can also impact phenomena at lower energies, \eg, flavor physics. However, the level of this impact depends upon 
the model details of both the generational dependence of the PM-SM mixings and the associated couplings to the new heavy gauge bosons. Due to this significant model dependence (that we 
have so far avoided), we will not consider these interesting possibilities here.

The most obvious place to begin is with the Drell-Yan production of the $Z'_M$ in $\bar qq$ annihilation at hadron colliders, \eg, the LHC and the 100 TeV FCC-hh, as this state can couple to a purely 
SM initial state, be resonantly produced on-shell, and decay into leptons pairs with a significant branching fraction.  In the narrow width approximation (NWA) the signal rate is proportional to
the $Z'_M$ resonant production cross section times its leptonic branching fraction, $B_l$. For the SM fields, which we recall all have $Q_D=0$, we can write the relevant couplings in terms of 
the parameters ($a,b$) that were introduced previously [which we recall are functions of both $r$ (and so also $\sigma$) and $\kappa_L$] as just 
\begin{equation}
g_L\Big(a~\frac{T_{8L}}{\sqrt 3}+b~\frac{Y}{2}+\frac{2}{3}a\kappa_L^2~Q_D\Big) Z'_M\\,
\end{equation}
with the last term just being zero in the case of the SM fields. We also recall from above, for numerical purposes, that for $\sigma = \pm1$, $b$ is relatively small in comparison to $a$ so that 
we expect reduced sensitivity to the {\it sign} of $\sigma$ in production cross sections. Since the values of both $T_{8L}$ and $Y/2$ are fixed for the SM fields, all of the relevant couplings are 
known quantities apart from the values of $\kappa_L$ and $\sigma$. For later usage, we note that the partial width for the $Z'_M$ decay into (massless) left-handed SM leptons (which we recall have 
$T_{8L}/\sqrt 3=1/6$) is given, for simplicity, in the approximate $|b/a| \to 0$ limit by
\begin{equation}
\Gamma_|=\Gamma(Z'_M \to l^+l^-) \simeq \frac{g_L^2M_{Z'_M}}{864\pi}~a^2\\.
\end{equation}
Further, apart from the overall QCD/color factors, \ie, $N_c=3$, and small QCD corrections, $\simeq (1+\alpha_s(M_{Z'_M})/\pi)$, that appear for the quarks, all of the SM fermions will essentially 
have this same partial width in the $|b/a| \to 0$ limit when the top mass can be safely   
neglected. Observe that when $a^2$ gets sufficiently large near the upper limit of the allowed range for $\kappa_L$ we will loose perturbativity and we so can no longer be able to trust our results 
and certainly not the NWA to be employed below. Clearly, the largest that $B_l$ can be is when the $Z'_M$ decays {\it only} into just the SM fermion final states as any 
additional modes would increase its total decay width. If this condition is satisfied, then we find that $B_l \simeq 0.04$ but, more generally, $B_l$ is instead 
\begin{equation}
B_l=\frac{\Gamma_l}{\Gamma_T+{\cal N}\Gamma_l}\simeq \frac{1}{25+{\cal N}}\\,
\end{equation}
where we've parameterized any additional contributions to the total $Z'_M$ decay width via the quantity ${\cal N}=\Gamma(Z'_M \to {\rm new})/\Gamma(Z'_M \to l^+l^-)$ which is zero in the simplest 
case with $\Gamma_T$ being the total width in this case. Thus we will be interested in knowing how any additional partial decay widths will scale with respect to that for the SM leptons given above 
in what follows. Certainly, if ${\cal N}$ becomes very large 
the NWA approximation will also fail as the $Z'_M$ total width to mass ratio will be become too large - as may also happen when $a$ becomes too large even without the introduction of any new 
additional decay modes as already noted.  Below we frequently employ this simplifying assumption, \ie, the absence of any additional new physics, so that if $|b/a|<<1$ then we will have 
$B_l \simeq 4\%$ as previously noted.  Any bound on ${\cal N}$ will be much more easily saturated when $\sigma =\pm 3$ than if $\sigma =\pm 1$ as much larger values of $a^2$ are obtained 
in that case.

Fig.~\ref{figb} shows the rate for $Z'_M$ production as a function of its mass under the assumption that only SM fermion final states can appear for either choice of $\sigma=\pm 1$ and 
with different selections of the value of $\kappa_L$ at both the LHC and the 100 TeV FCC-hh.  Here we see that, as expected, this rate is relatively insensitive to $b$ and, hence, the sign of 
$\sigma$. We also see that it grows as $\kappa_L$ approaches its maximum value, $1/\sqrt r \simeq 0.821$, due to the factor of $(1-\kappa_L^2 r)^{1/2}$ appearing in the denominator of 
the definition of $a$. When $\sigma=\pm 3$, for a given value of $\kappa \lsim 0.269$, both $r$ and $a$ are now larger which would then result in an enhanced production cross section and a 
potentially increased influence of the sign of $\sigma$ as $r$ would be somewhat larger in such a case. In practice, however, we still find that this sign will have little practical influence on 
the search reaches.

Both ATLAS\cite{ATLAS:2019erb} and CMS\cite{CMS:2021ctt} have performed $Z'$ searches in the dilepton channel at the 13 TeV LHC employing an integrated luminosity of 139 fb$^{-1}$ which 
we can apply to the case at hand; here we will employ the specific results obtained by ATLAS. Similarly, we can follow the analysis as presented in Ref.\cite{Helsens:2019bfw} to obtain the 
corresponding expected reach for the 100 TeV FCC-hh assuming 30 ab$^{-1}$ of integrated luminosity. The results of these analyses are shown in Fig.~\ref{figc} as functions of $\kappa_L$ 
assuming that $\sigma =\pm 1${\footnote {As we will see below, when $\sigma= \pm 3$, these cross sections will we seen to increase by an order of magnitude or more which results in an 
increased mass reach of $\simeq 20-25\%$}}.  Here we see that the bounds at the 13 TeV LHC  
are relatively modest and increase slowly at first as $\kappa$ increases away from zero with the choice of $\sigma=1$ yielding the slightly 
larger values; as a point of comparison, a heavy SM-like $Z'$, $Z'_{SSM}$, is constrained to lie above $\simeq 5.1$ TeV from this search\cite{ATLAS:2019erb}. However, again due to the factor of 
$(1-\kappa_L^2 r)^{1/2}$, when $\kappa_L$ exceeds roughly $\simeq 0.64$, the limit strengthens significantly due to the now much more rapid growth of the parameter $a$. Assuming no signals 
are found, the 14 TeV HL-LHC with a luminosity of 3 ab$^{-1}$ is expected to increase these bounds by roughly $\sim 10-15\%$. The corresponding results for the 100 TeV FCC-hh in the lower 
panel of the Figure show quite similar behavior; in this case we note that the $Z'_{SSM}$ bound is determined to be $\simeq 42$ TeV from Ref.\cite{Helsens:2019bfw} for comparison purposes. 

\begin{figure}[htbp]
\centerline{\includegraphics[width=5.0in,angle=0]{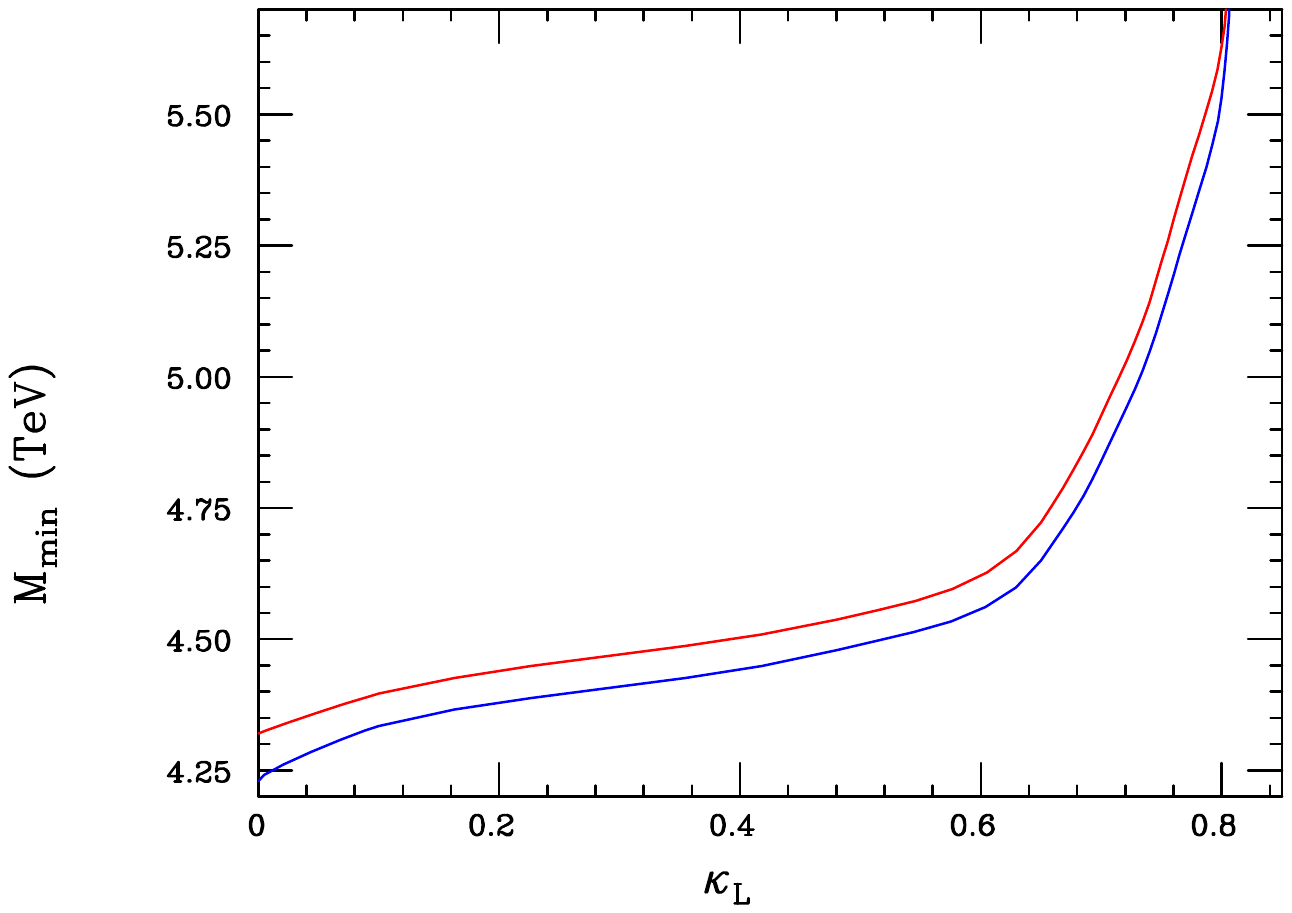}}
\vspace*{-0.8cm}
\centerline{\includegraphics[width=5.0in,angle=0]{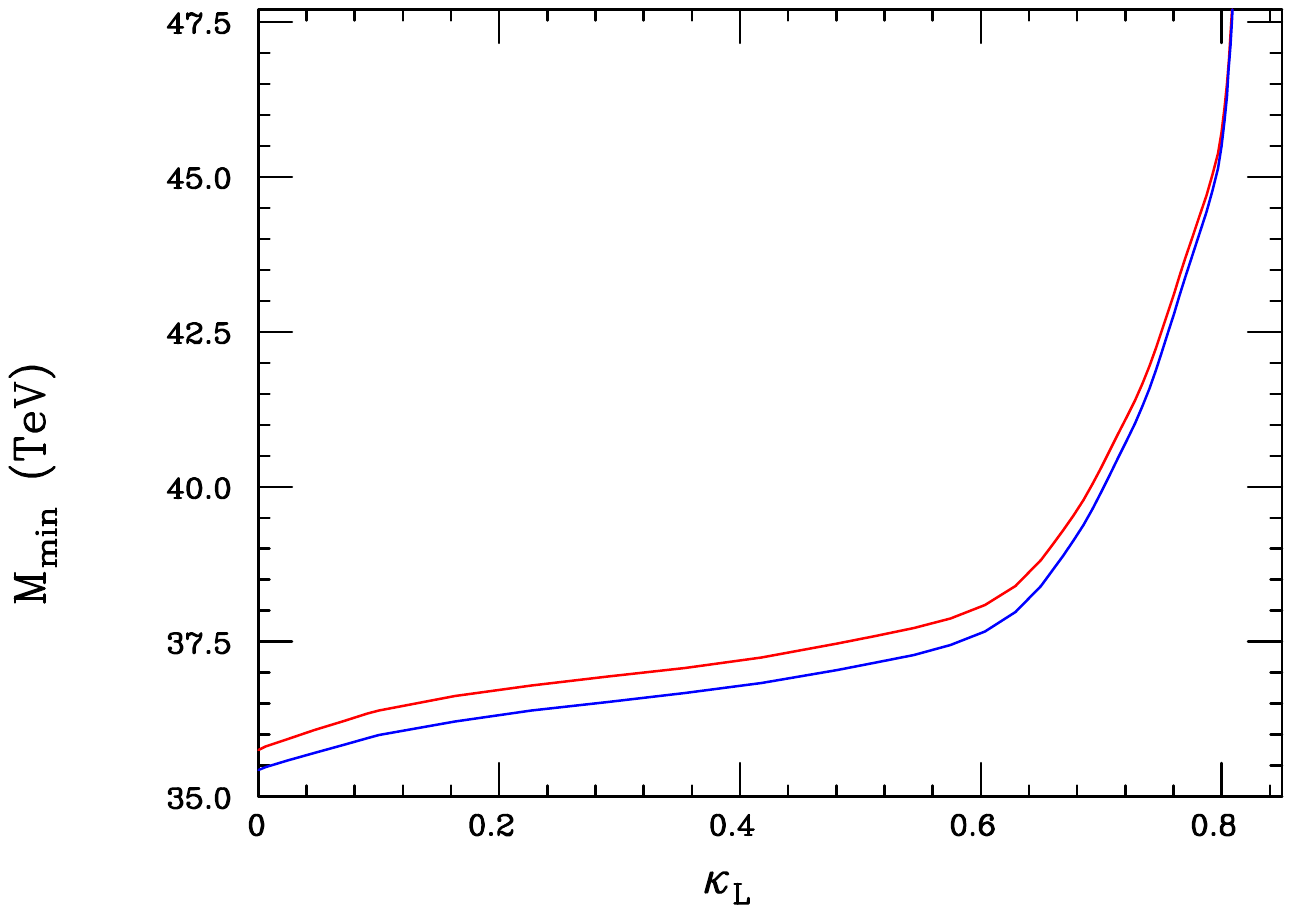}}
\vspace*{-1.3cm}
\caption{$Z'_M$ mass bounds as functions of $\kappa_L$ with $\sigma=1(-1)$ corresponding to the red (blue) curve, following the analysis described in the text for (Top) the 13 TeV LHC 
employing the results from ATLAS\cite{ATLAS:2019erb} and (Bottom) for the 100 TeV FCC-hh assuming an integrated luminosity of 30 ab$^{-1}$ and employing the analysis as presented 
in Ref.\cite{Helsens:2019bfw}, respectively. The NWA approximation is employed in obtaining these results and assume decays only to the SM fermions.}
\label{figc}
\end{figure}

\begin{figure}[htbp]
\centerline{\includegraphics[width=5.0in,angle=0]{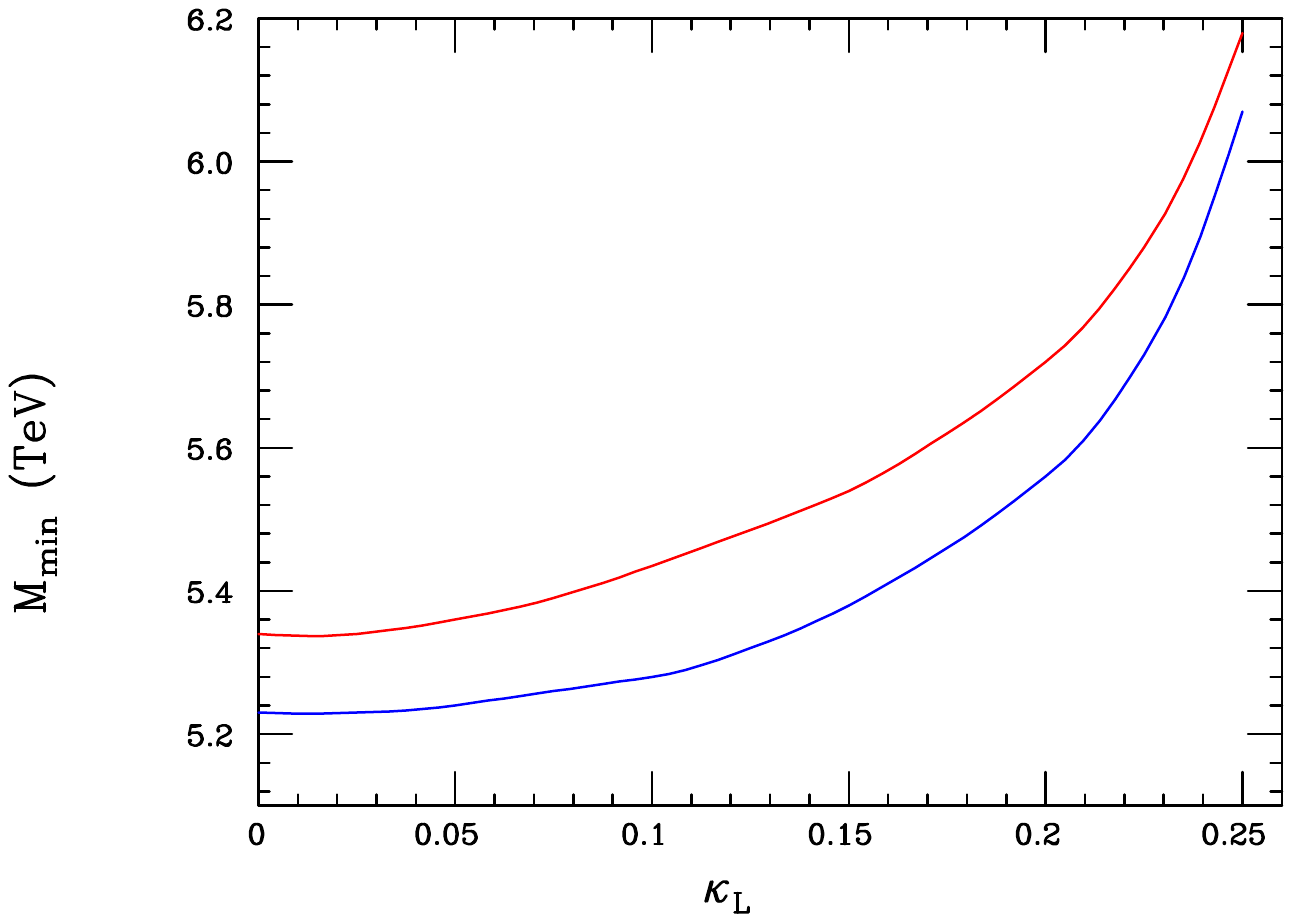}}
\vspace*{-0.8cm}
\centerline{\includegraphics[width=5.0in,angle=0]{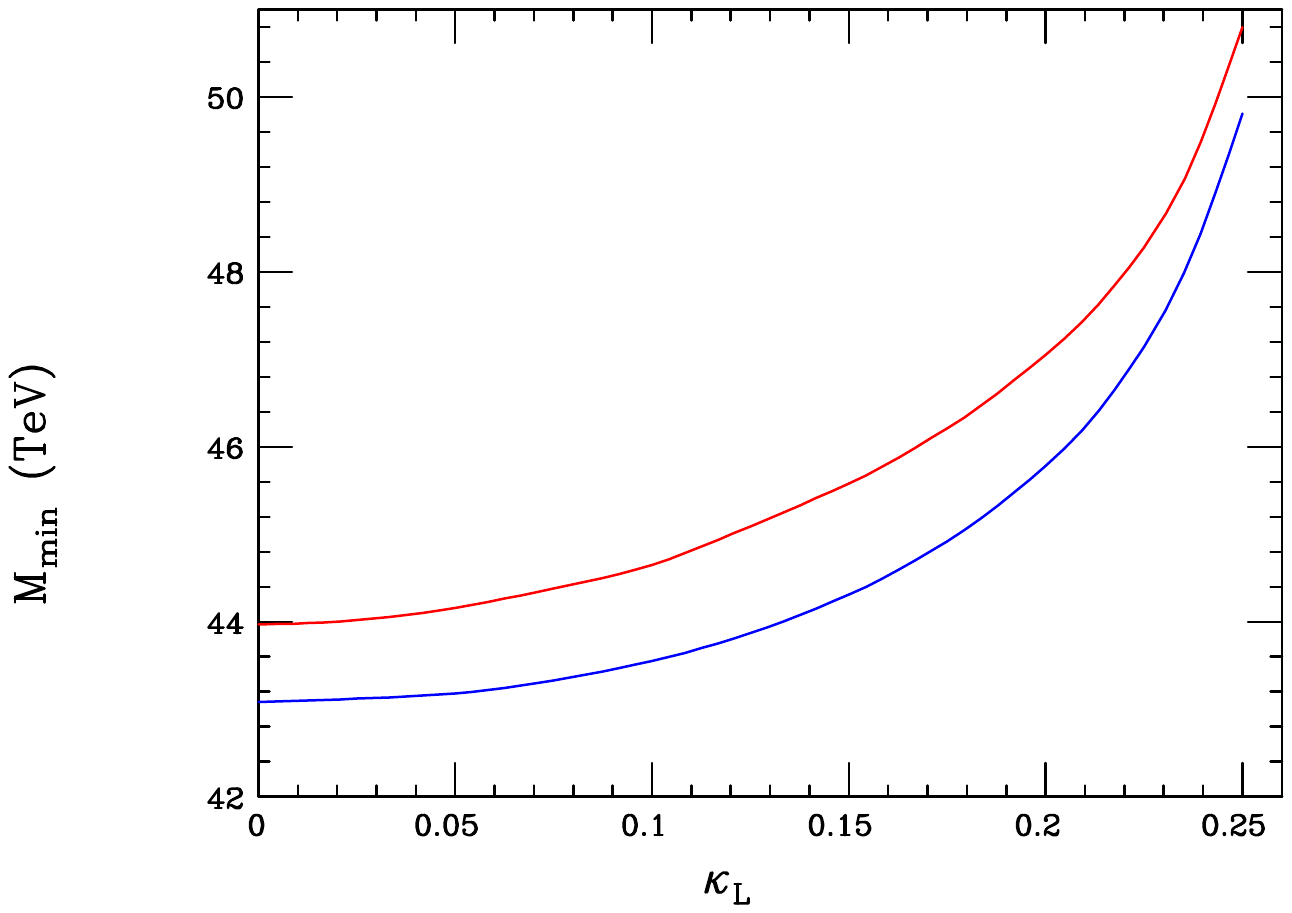}}
\vspace*{-1.3cm}
\caption{Same as the previous Figure, but now assuming that $\sigma=\pm 3$.}
\label{figc3}
\end{figure}

The corresponding bounds obtained in the case of $\sigma =\pm 3$ are shown in Fig.~\ref{figc3}. Here we see that, as expected, substantially greater exclusion and search reaches are obtained,  
by roughly $\simeq 20-25\%$, since the values of $a$ are now much larger. With the ratio of $|b/a|$ being somewhat larger in this case, as was noted above, we also observe a somewhat greater 
sensitivity to the sign of $\sigma$, also as expected.

\begin{figure}[htbp]
\centerline{\includegraphics[width=5.0in,angle=0]{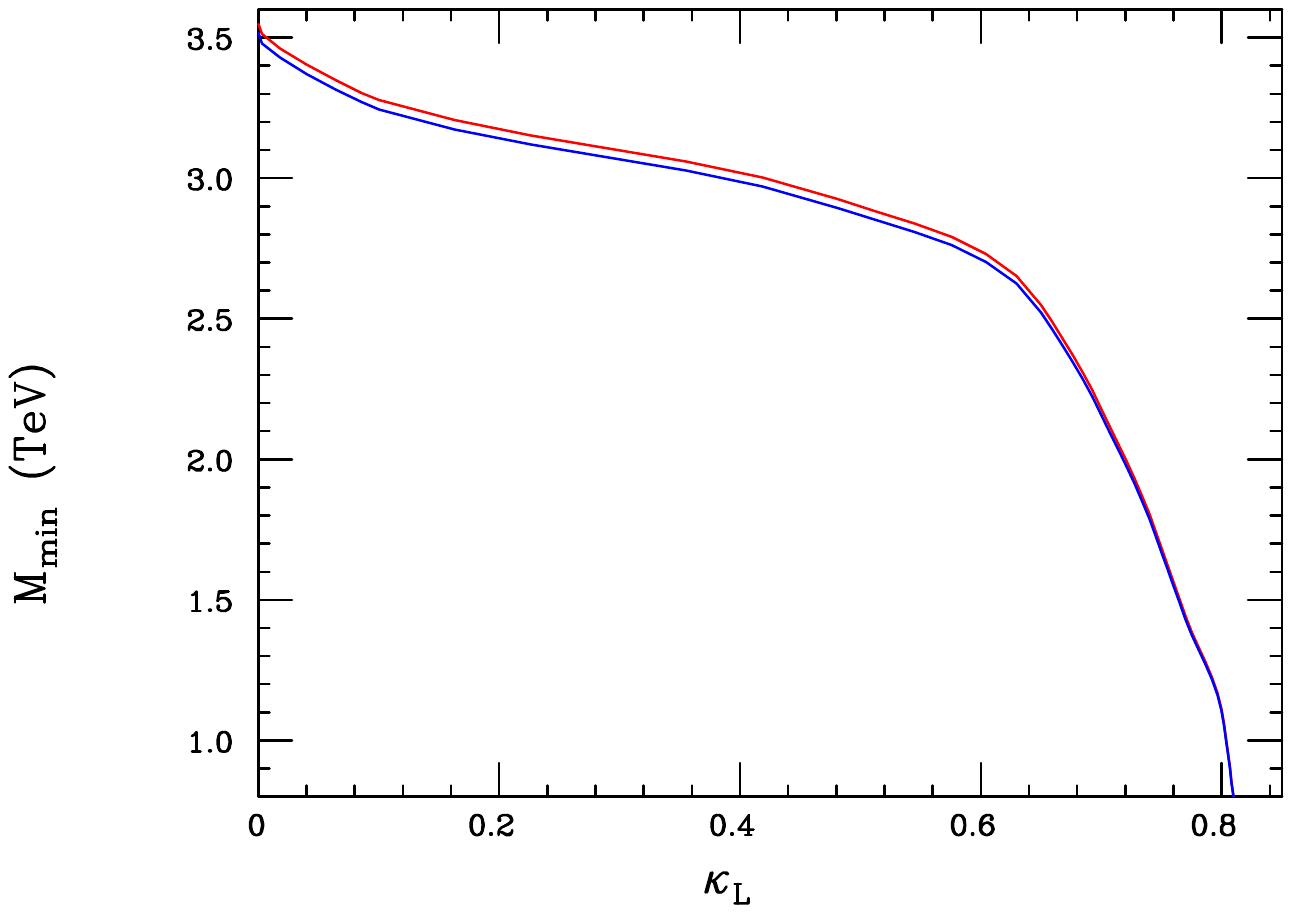}}
\vspace*{-0.8cm}
\centerline{\includegraphics[width=5.0in,angle=0]{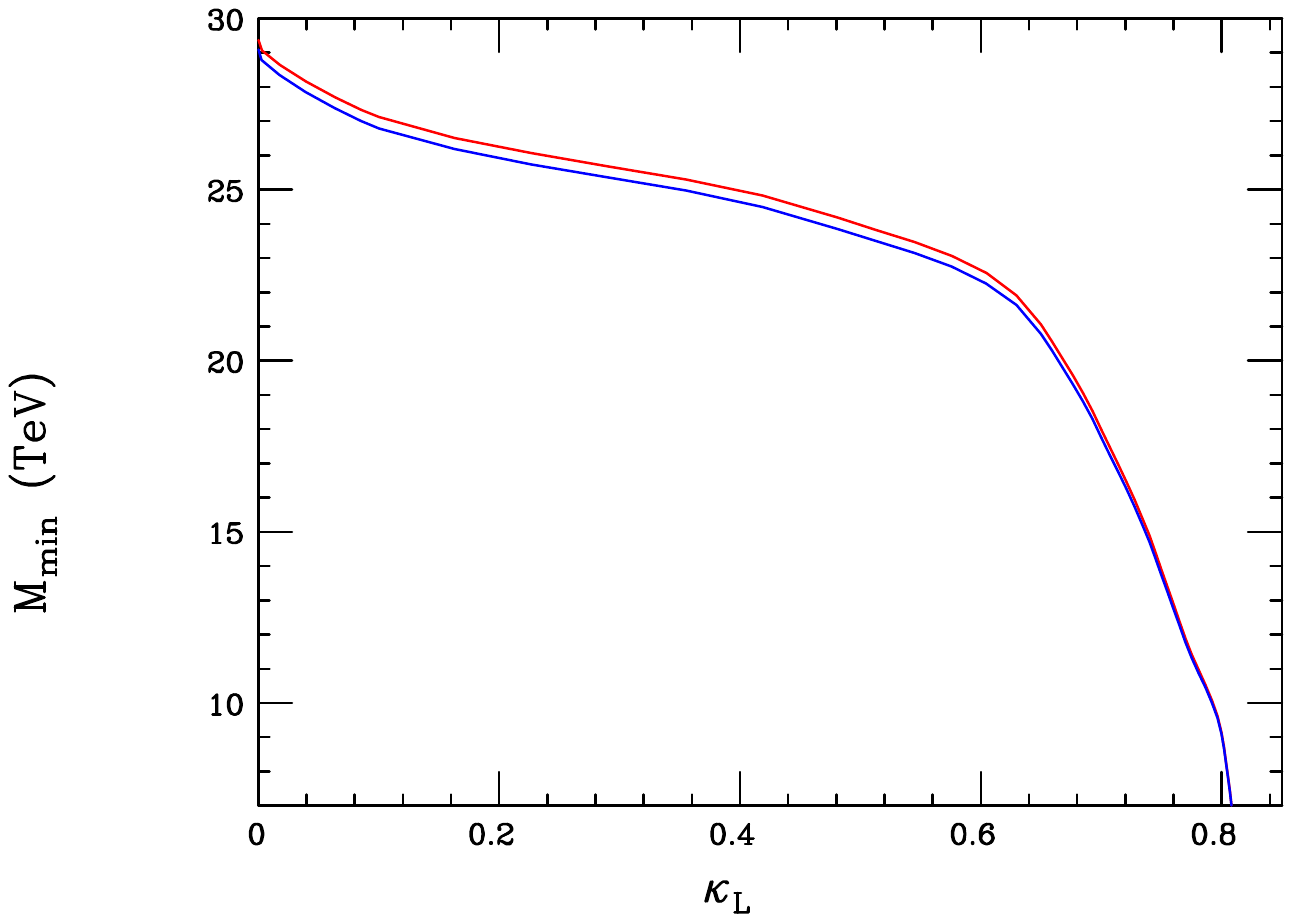}}
\vspace*{-1.3cm}
\caption{{\it Indirect} mass bounds on the NHGBs obtained by employing the model-dependent mass relationship given in the text and the $Z'_M$ mass constraints from the 
previous Figures here assuming that $\sigma =\pm 1$.  }
\label{figd}
\end{figure}

One simple and immediate application of the $Z'_M$ mass constraints obtained above is that, within the present model context, we can translate these bounds into the corresponding, 
yet {\it indirect}, constraints on the 
NHGB using the mass relationships discussed above. These are potentially important as they can, in some parameter space regions, supersede those from direct production to be discussed 
further below. In Fig.~\ref{figd} these constraints are shown, assuming $\sigma =\pm 1$, using the results as displayed in Figs.~\ref{figa} and ~\ref{figc} as input. These constraints are seen to 
be relatively strong for small values of $\kappa_L$ but then weaken substantially as $\kappa_L$ increases towards its upper bound as we might have expected. Correspondingly, indirect bounds are 
similarly obtainable for the case of $\sigma=\pm 3$ and display a comparable overall behavior. 

What about other $Z'_M$ decay modes beyond those to the SM fermions which may be significant? Such modes can exist even if we remain restricted to the familiar set of SM final states. Any such 
modes will lower $B_l$ and so lead to a (generally modest) reduction in the corresponding search reach using the Drell-Yan channel but they are also important to consider on their own.
Interestingly, as is well-known\cite{Hewett:1988xc}, the small mixing of a new gauge boson such as $Z'_M$ with the SM $Z$ can induce other decays modes which may have significant partial widths 
which can be comparable to those for that for the SM leptons. A classic example of this is the mode $Z'_M \to W^+W^-$ which is induced via this angle, $\theta_{mix}$, as encountered above in
Eqs.(44)-(46); in the present case, one finds that this partial width is given, to leading order in the small mass ratios, by 
\begin{equation}
\Gamma(Z'_M \to W^+W^-)\simeq \frac{g_L^2 M_{Z'_M}}{192\pi c_w^2}~~\Big(\theta_{mix} \frac{M_{Z'_M}^2}{M_Z^2}\Big)^2\\,
\end{equation}
where (using $M_W=M_Zc_w$ which remains true to leading order in the small vev ratios) we see the familiar result that the small mixing angle is offset by the square of the large ratio of the 
$Z'_M$ and SM $Z$. Using the expression above that relates this mixing angle to the other model parameters, $\theta_{mix}\simeq M_{int}^2/M_{Z'_M}^2$, we obtain from Eqs.(44)-(46) that
\begin{equation}
{\cal R}_W=\frac{B(Z'_M \to W^+W^-)}{B_l}\simeq \frac{1}{4} \cdot \Big[ \frac{v_2^2-v_1^2}{v_2^2+v_1^2}+\frac{3b}{a}\Big]^2\\,
\end{equation}
so that ${\cal R}_W$ may be appreciable $\simeq 0.25$ or so as the expression in brackets if $O(1)$. Clearly a relative branching fraction of this magnitude would have essentially no impact the 
searches above employing the dilepton channel. If the dilepton production rate is significant, then the $W^+W^-$ mode can also be sufficiently large to be used to probe this gauge boson mixing 
and the symmetry breaking structure of the model. The cleanest signature for the observation of this decay, since the $W$'s are expected to be highly boosted, would likely be an almost collinear 
lepton + MET combination plus two collimated jets in the opposite hemisphere with both systems reconstructing to the $W$ mass. 

\begin{figure}
\centerline{\includegraphics[width=5.0in,angle=0]{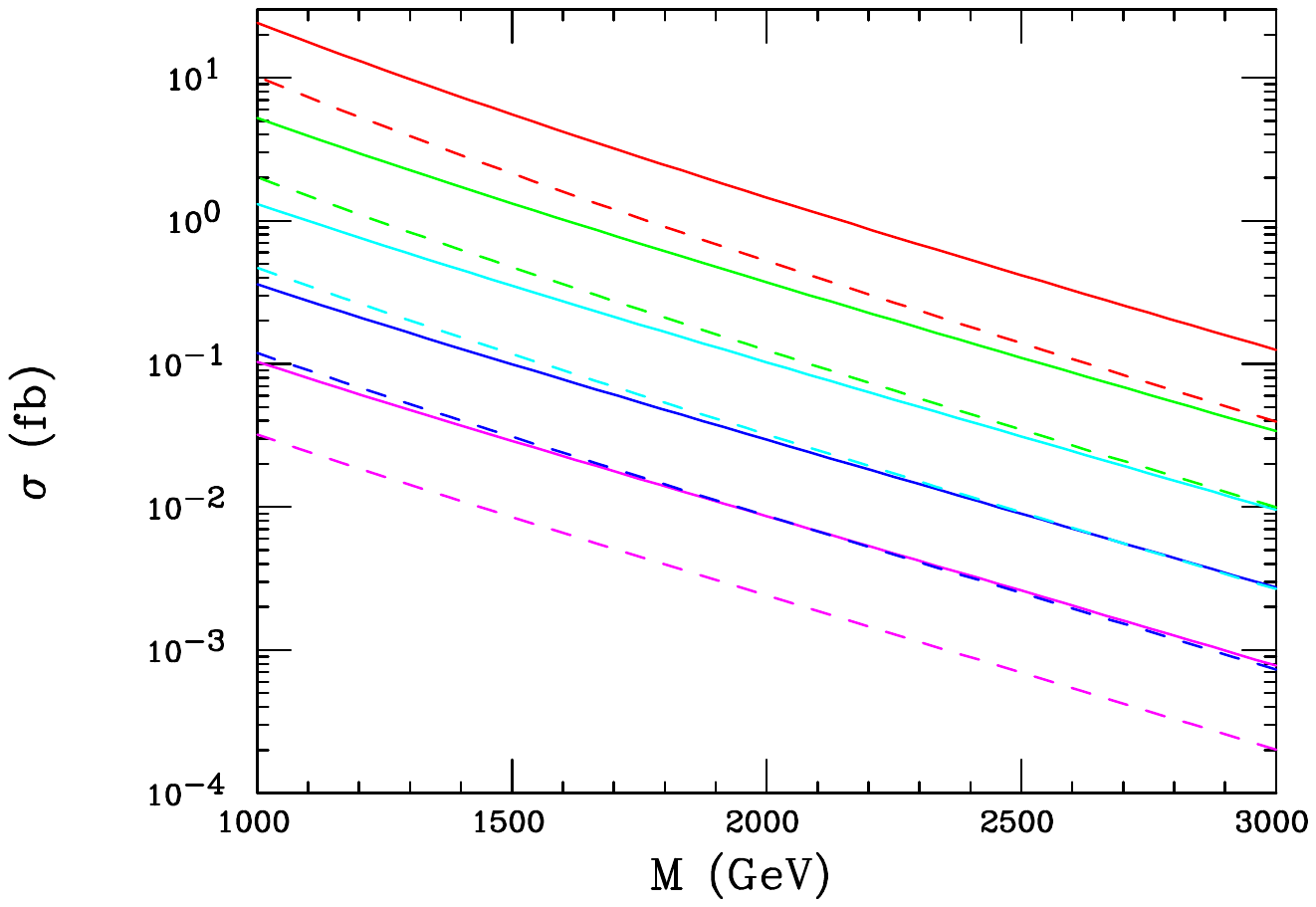}}
\vspace*{-2.3cm}
\centerline{\includegraphics[width=5.0in,angle=0]{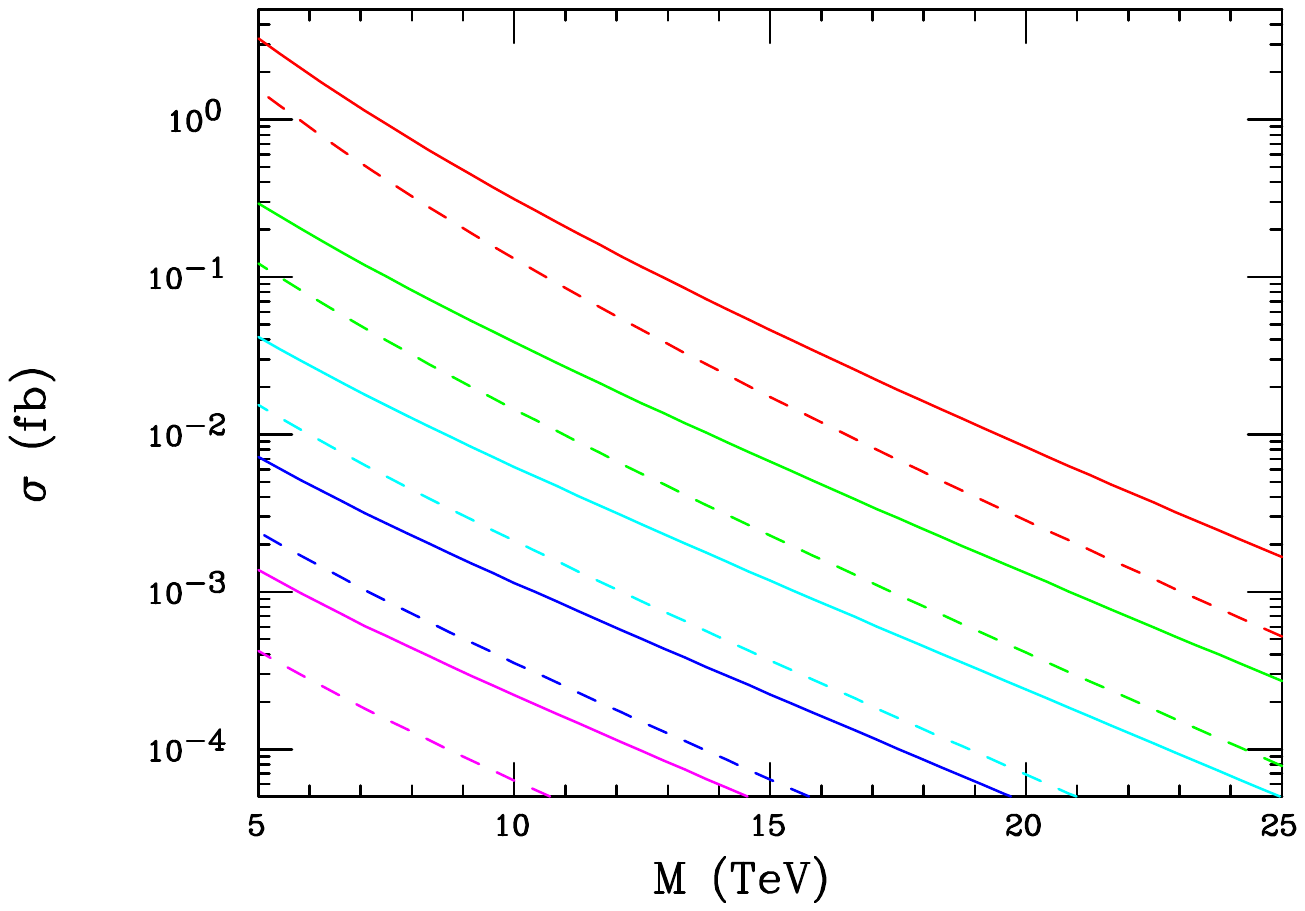}}
\vspace*{-1.30cm}
\caption{$gu(gd)$-initiated NHGB plus PM associated production cross section represented by the solid (dashed) curves at the (Top) 13 TeV LHC and (Bottom) 100 TeV FCC-hh as functions of 
the appropriate NHGB mass. In the Top panel, from top to bottom, the curves are for a PM mass of 1, 1.5, 2, 2.5 and 3 TeV, respectfully, whereas in the Bottom panel, the corresponding curves 
label the results for PM masses of 5, 10, 15, 20 and 25 TeV, respectively.}
\label{fige}
\end{figure}

The NHGB, which we recall carry $Q_D\neq 0$,  can be produced at hadron colliders via two mechanisms: generically, the first of these, associated production, can be described by the 
$gq(\bar q) \to {\rm NHGB} +{\rm PM}$ process where $g$ is a gluon and $q(\bar q)$ is a either a SM $u$ or $d$-type (anti)quark - not dissimilar to a previously examined production process 
studied for both the LHC and FCC-hh\cite{Rizzo:2023kvy}.  Depending upon 
whether the quarks lie in a $q_1$ or a $q_2^*$ representation, as well as the value of $\sigma$, several distinct channels are possible, \eg, $gu\to B^+D,A^0U$ in the language employed above. 
Since the overall coupling, $g_L$, is fixed, only the PM and NHGB masses are {\it a priori} unknown. Fig.~\ref{fige} shows the relevant cross sections for both the $gu$ and $gd$-initiated processes 
at the 13 TeV LHC and at the 100 TeV FCC-hh as functions of the masses of these two heavy particles in the final state. Certainly the signatures for this production process will depend somewhat 
on the model details such as the mass ordering of the PM and NHGB, neither of which are not likely to be very boosted since they are expected to be quite massive. For example, in the simple 
scenario where the NHGB is the heavier of the two, then we can have decays such as ${\rm NHGB}\to {\rm PM}+\bar q$ while ${\rm PM}\to qV$, which leads to a $3j+$MET final state with the 
MET coming 
from the two DP in the final state which are assumed to be either long-lived or which decay into DM. Many other similarly interesting modes can occur as any given model will have both color triplet 
as well as color singlet PM field content. For example, the NHGB may instead dominantly decay to a leptonic $\bar eE$ pair followed by the $E\to eV$ leading to a $1j$+ a 
non-resonant, opposite sign dilepton pair +MET final state. By combining several such modes a substantial discovery reach is very likely obtainable. From Fig.~\ref{fige}, we see that regions of 
parameter space corresponding 
to the sum of the NHGB and PM masses up to (very) roughly $\simeq 4.0(3.5)$ TeV and $\simeq 27(25)$ TeV for the $gu(d)$-initiated process may be obtainable at these two colliders, respectively.

\begin{figure}[htbp]
\centerline{\includegraphics[width=5.0in,angle=0]{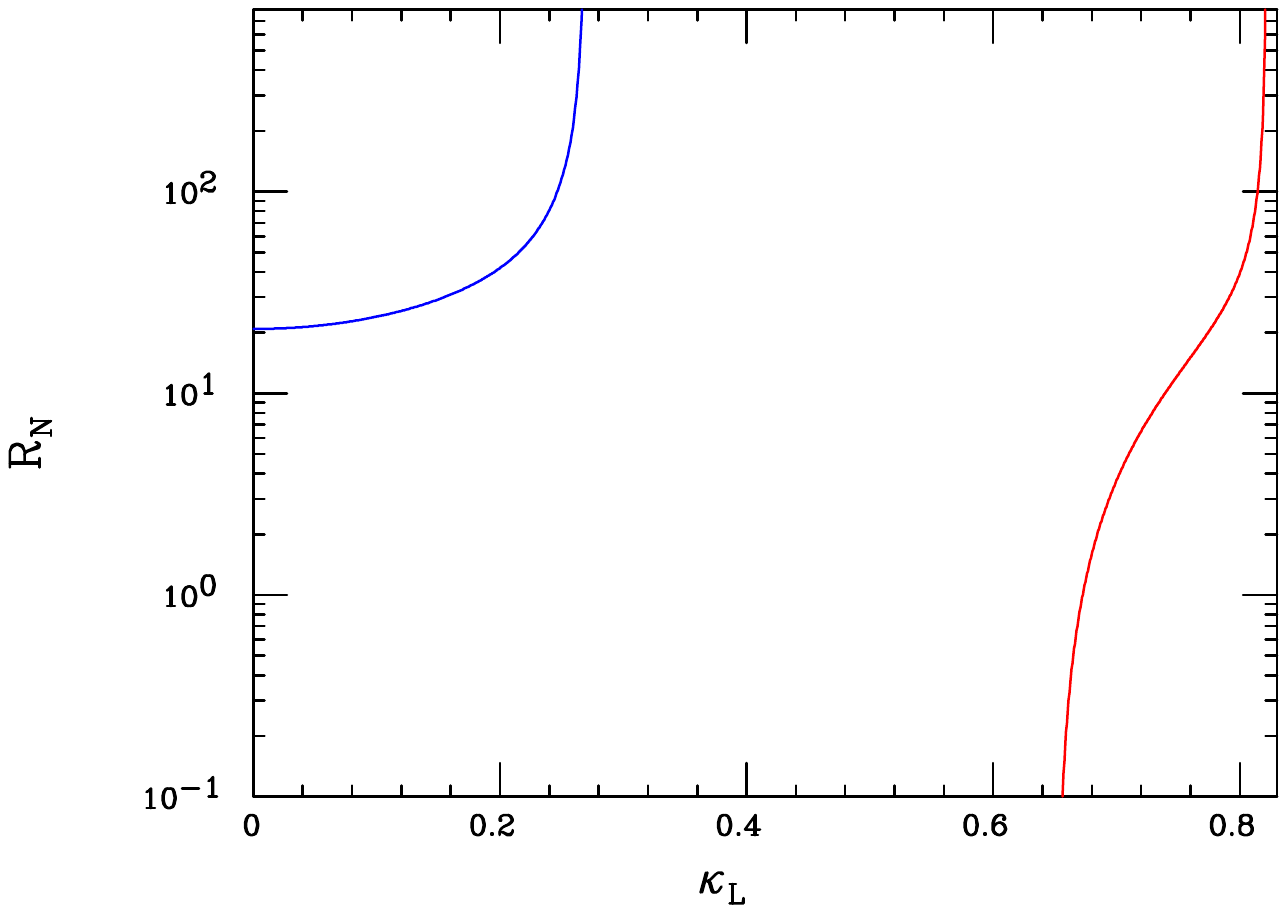}}
\vspace*{-1.3cm}
\caption{The ratio, ${\cal R}_N$, described in the text as a function of the parameter $\kappa_L$, for both $|\sigma| =1$ (red) and $|\sigma|=3$ (blue).  Recall that in both cases the allowed 
range of $\kappa$ is restricted from above.}
\label{figz}
\end{figure}

The second mechanism for making these NHGB is pair production via $s$-channel exchanges, \ie, $q\bar q \to \gamma, Z, Z'_M \to {\rm NHGB}+ {\rm NHGB}^\dagger$, where the SM $\gamma$ 
exchange may or may not 
contribute depending upon the particular model details and the specific NHGB under consideration. Recall that both of the NHGB carry weak isospin having $T_{3L}=\pm 1/2$, and at least one 
of them has a non-zero electric charge. Since, unlike in the case of associated production, pair production is a purely electroweak process the 
cross section in this case is generally small. Importantly, however, the rate for this reaction can also be potentially $Z'_M$ resonance enhanced as was discussed earlier provided that 
$M_{Z'_M}> 2M_{NHGB}$. As we saw above, when $|\sigma|=1$, one finds that $\kappa_L$ is restricted to a narrow range to allow for this possibility whereas for, $|\sigma|=3$, there is no such 
restriction. It thus behooves us to examine the ratio of branching fractions ${\cal R}_N=B(Z'_M \to {\rm NHGB}+{\rm NHGB}^\dagger)/B_l$ above this threshold, which is similar to that 
for the $W^+W^-$ final state above except that there is no longer any small mixing angle suppression and that the $Z'_M$ and NHGB masses are now more than likely be comparable. Quite 
generally we may write (again for simplicity assuming that we may approximately take $|b/a|\to 0$ as above) 
\begin{equation}
\Gamma(Z'_M \to {\rm NHGB}+{\rm NHGB}^\dagger)= \frac{g_L^2 M_{Z'_M}}{192\pi}~ P^2~\frac{1}{X^2}\big(1-4X\big)^{3/2}\big(1+20X+12X^2\big)\\,
\end{equation}
where, from the above discussion and Eq.(42) one obtains in the present setup that 
\begin{equation}
X=\frac{M_{NHGB}^2}{M_{Z'_M}^2}=\frac{1-\kappa_L^2r}{r}\leq \frac{1}{4}\\,
\end{equation}
and we find, employing Eqs.(14) and (43) as well as the subsequent discussion, that the parameter $P^2=9/(4a^2)=X$. Thus we finally arrive at the required expression: 
\begin{equation}
{\cal R}_N=\frac{1}{X} \big(1-4X\big)^{3/2} \big(1+20X+12X^2\big)\\,
\end{equation}
which we see can very easily be larger than unity when $X$ becomes small. For example, taking $|\sigma|=1$ and $\kappa_L=0.7$, one finds that ${\cal R}_N \simeq 3.7$ for {\it each} of the NHGB 
which is quite substantial; when $|\sigma|=3$, even larger values of this ratio are obtained.  Fig.~\ref{figz} shows the values of ${\cal R}_N$ as a function of $\kappa_L$ when $|\sigma|=1$ or 3; the 
reader is reminded of the restricted ranges of $\kappa_L$ in these two case as discussed above when examining the results as presented in this Figure. 

On the positive side, this enhancement is quite advantageous as it allows us to study the production of the NHGB up to quite large masses assuming that they can be pair-produced on-shell. 
Simultaneously, however, the relatively large partial width for this process begins to significantly reduce the value of $B_l$ which will negatively impact the discovery reach for 
the $Z'_M$ in the dliepton channel when very large values of ${\cal R}_N$ are realized. However, even when rather sizable values of this ratio are indeed obtained, \eg, 25-50, the actual search 
reach is found to be not very much degraded, roughly by less than $\sim 5-10\%$,  in practice.

\begin{figure}[htbp]
\centerline{\includegraphics[width=5.0in,angle=0]{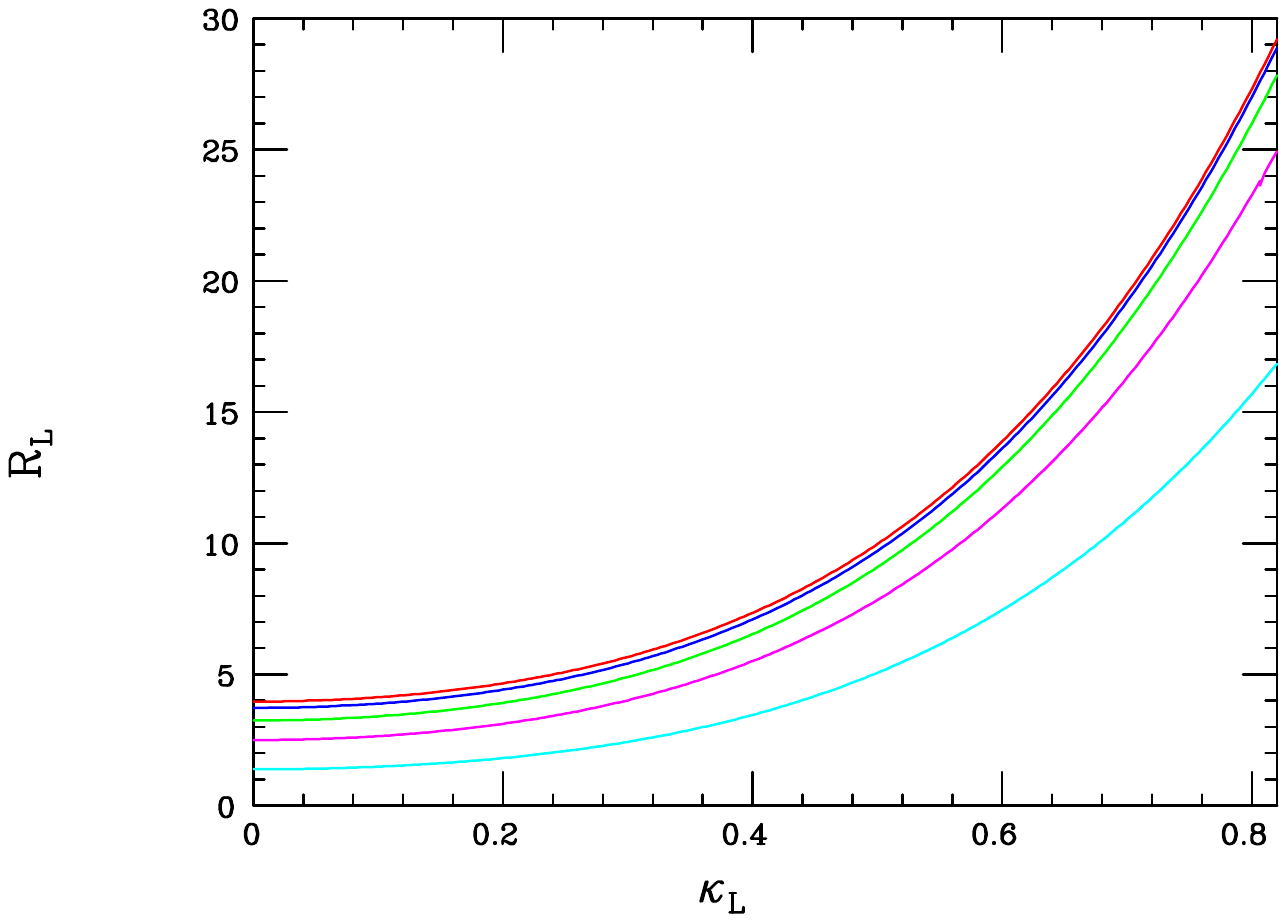}}
\vspace*{-0.8cm}
\centerline{\includegraphics[width=5.0in,angle=0]{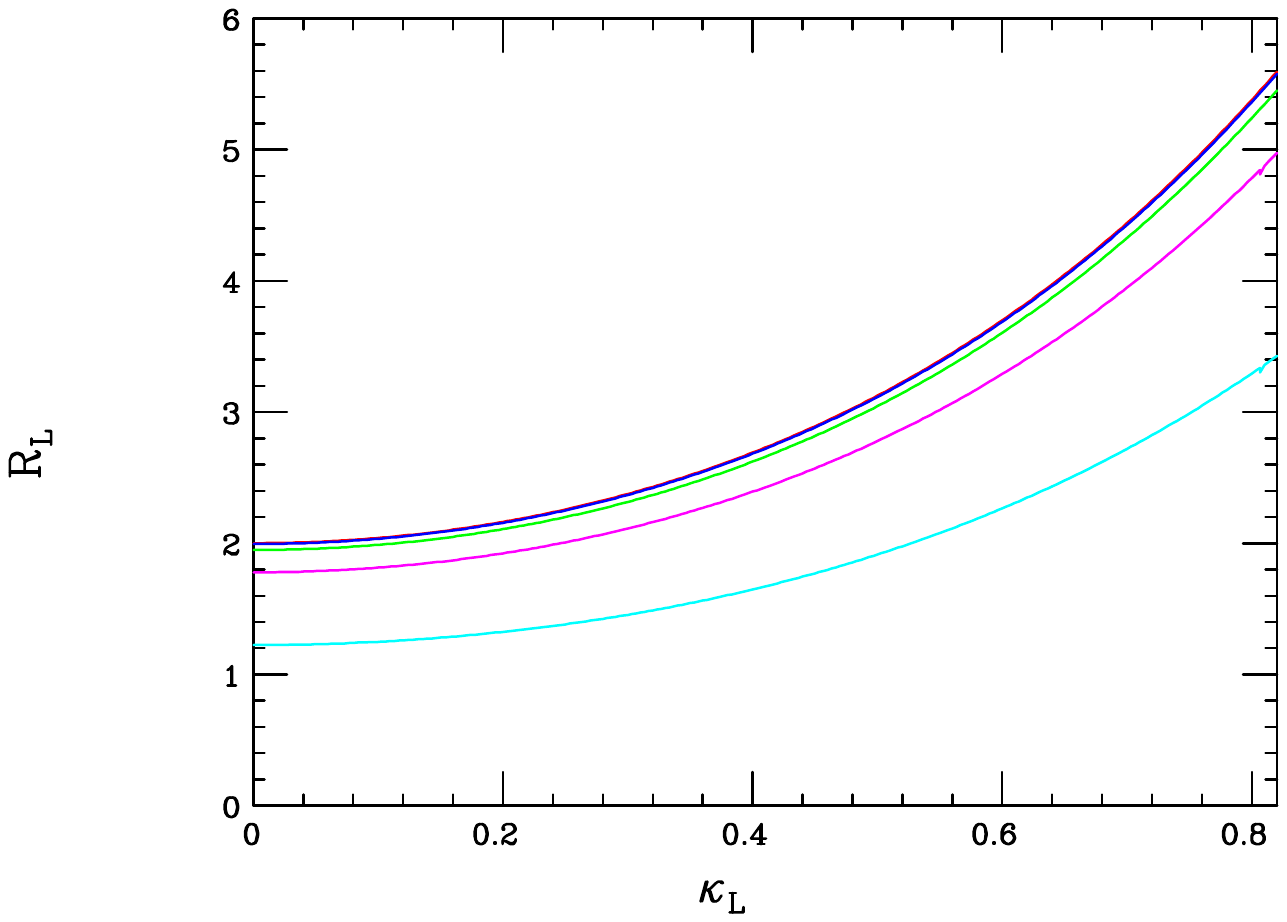}}
\vspace*{-1.3cm}
\caption{The ratio, ${\cal R}_L$, of the leptonic PM $Z'_M$ branching fraction to that for ordinary SM charged leptons, described in the text, as a function of the parameter $\kappa_L$ assuming, 
from top to bottom, $m/M_{Z'_M}=0.05$ (red), 0.15 (blue), 0.25 (green), 0.35 (magenta) and 0.45 (cyan), respectively, where $m$ is the $N$ or $E$ PM mass in the case where such fields are 
(Top) right-handed singlets or are (Bottom) vector-like with respect to $3_L$ as discussed in the text. }
\label{figf}
\end{figure}

Turning now to the heavy PM fields, in the case of the $U-$ or $D-$type color triplets, these states can be pair produced though QCD in analogy to the top quark in the SM via $gg$ and $\bar qq$. 
As mentioned  previously, these particles will dominantly decay to the analogous SM states with which they share a ${\bf 3}$ or ${\bf{\bar 3}}$ multiplet together with a DP so that the simplest final 
state (for non-$b$ or $t$ quarks) is just $2j+$MET, similar to that of squark production. As discussed in earlier work\cite{Rizzo:2022qan}, these types of searches at the 13 TeV LHC already exclude 
$U/D$ 
masses below $\simeq 1.5-1.8$ TeV, depending upon the flavor of the SM quarks in the final state, but at the FCC-hh the corresponding search reaches increases substantially to roughly 
$\simeq 10$ TeV. Analogously, the heavy PM fields which are color singlets, $N$ and $E$, can only be produced by electroweak exchanges, \ie, $\bar q q\to \gamma, Z, Z'_M \to \bar NN, E^+E^-$ 
and so their cross sections are relatively small. In the case of $E$, LHC searches employing the assumed dominant $eV$ (or $eh_D$) decay mode and producing an opposite sign lepton plus MET 
signature already exclude PM masses up to roughly $\simeq 1$ TeV\cite{Rizzo:2022qan}. 
However, if these PM leptons can appear in the decays of $Z'_M$, as was the case with the NHGB, we will have a resonance enhanced search reach. In the case of $|\sigma|=1$ where the 
anomalies cancel among the 3 generations, these PM leptons, $N,E$, will lie in either $l_1$- or $l_2^*$-like triplet/anti-triplet representations in a manner similar to that of $U,D$ just discussed. 
In such a case, where the right-handed components of these PM leptons are singlets of $3_L$,  the relevant quantity of interest is then just the ratio 
${\cal R}_L=B(Z'_M \to \bar NN ~{\rm or}~ E^+E^-)/B_l$ which, where if we again make the now familiar approximation $|b/ a| \to 0$ and recall that $|Q_D(N,E)|=1$, is given by an identical expression 
for both $N,E$ types of PM leptons: 
\begin{equation}
{\cal R}_L=2\beta ~\Big[\beta^2+ \frac{3-\beta^2}{2} (1+4\kappa_L^2)^2 \Big]\\,
\end{equation}
with $\beta^2=1-4m^2/M_{Z'_M}^2$ and where $m$ is now either $m_{N,E}$. As the top panel of Fig.~\ref{figf} shows, ${\cal R}_L$ can be quite large, even near the mass threshold, when 
$\kappa_L$ becomes large. A similar expression would hold if $Z'_M$ were allowed to decay into pairs of the color triplet PM fields, apart from an overall factor of $\simeq 3(1+\alpha_s(M_{Z'_M})/\pi)$, 
which under some circumstances may match or exceed the rate that is expected from the pure QCD production process already mentioned, especially at the 100 TeV FCC-hh. This could push 
the discovery reach for these states up to roughly the kinematic limit $\simeq 0.5 m_{Z'_M}$, similar to that for the color singlet states.

If the vector-like (with respect to $3_L$) leptons of the type appearing in $S_{9,10}$ (\ie, $N_2,E$ in $S_9$ or $N_1,E_2$ in $S_{10}$, respectively) are also/instead of  
encountered, then the corresponding cross sections are somewhat reduced in comparison to previous result but can still be quite significant and we would now obtain
\begin{equation}
{\cal R}_L=2\beta ~\Big[\frac{3-\beta^2}{2}\Big] (1+\kappa_L^2)^2 \\,
\end{equation}
as is shown in the bottom panel of this same Figure. These models with the anomalies cancelling within each generation will potentially also lead to more complex signatures in the leptonic 
sector due to cascading decay effects mentioned above especially if a significant fraction of the color singlet states are kinematically accessible in $Z'_M$ decays.

Clearly the production of the color singlet and triplet PM states as well as the new heavy gauge bosons associated with the enlarged gauge symmetries can lead to some interesting collider 
phenomenology.

\section{Discussion and Conclusion}

The kinetic mixing between the dark photon, arising from a $U(1)_D$ gauge symmetry, and the SM photon to mediate the interactions between a light DM and ordinary particles provides a simple 
yet elegant portal to explain the observed relic density with potentially numerous phenomenological implications. However, KM necessitates the existence of a set of fields which carry both SM 
and dark sector quantum numbers, \ie, portal matter, which must consist of scalars and/or fermions which are vector-like with respect to their SM interactions due to numerous unitarity, Higgs and 
precision measurement constraints.  The combination of the constraints arising from the CMB and the requirement that the coupling of the $U(1)_D$ gauge group associated with the DP remain 
perturbative up to some large scale, $\sim 10$'s of TeV,  possibly hints at the existence of some enlarged non-abelian group, $G_D$. When this gets broken down to (at least) $U(1)_D$, it provides 
the masses for the PM fields as well as to a set of new heavy gauge bosons, some of which will carry dark quantum numbers. It is then natural to ask how these ideas can fit together with the SM in 
a more unified framework. Is there a way to understand how the PM fields carrying both SM and dark quantum numbers arise in such a setup, what their properties might be, and how they and the 
new heavy gauge bosons might be directly discovered at colliders? 

In our previously considered bottom-up approaches that attempted to address these and related questions, we examined an enlarged gauge group, $G$, essentially consisting of a product of the 
SM and dark gauge groups, wherein, \eg, the $SU(2)_L$ part of the SM and $U(1)_D$ were totally uncorrelated and independent. One might easily imagine that this would no longer be true in a 
more fully unified picture so it behooves to examine alternative possibilities. In the present work, following a parallel bottom-up approach, we have explored a (relatively) simple toy setup wherein 
the $SU(2)_L$ part of the SM gauge group as well as part of $U(1)_D$ gauge group associated with the DP are unified within a single $SU(3)_L$ factor. This $SU(3)_L$ is also accompanied by 
two $U(1)$ factors, $1_A1_B$, thus providing for partial unification of the dark and SM electroweak gauge interactions. Explicitly, in such a setup the PM fields naturally (mostly) lie in common 
representations of $3_L$ along with the corresponding SM fields having similar quantum numbers. In this setup, the Higgs fields responsible for the breaking $3_L1_A1_B$ down to the 
SM$\times U(1)_D$ are directly responsible for generating the PM masses whereas the Higgs vevs which break $1_D$ also lead to a mixing of the PM fermions with analog SM fields, thus 
providing them with a likely dominant decay path and a unique collider signature. This is perhaps the simplest scenario which is not a product group of the form $G_D\times G_{SM}$ which allows 
for a (at least partial) unification of the SM electroweak with the  dark sector. If we consider the breaking $3_L\to 2_L1_L$, with $2_L$ identified as the usual SM group in this setup, then the SM 
hypercharge, $1_Y$, is embedded as a linear combination of the two factors $1_L1_A$ while $1_D$ is instead embedded as an orthogonal combination of all three factors, $1_L1_A1_B$.

Although several variants of this general setup are possible, depending upon the assumed value of the parameter $\sigma$ and the manner in which the anomalies are cancelled, they all have most 
of their features in common, which we take advantage of in the current study. First, given the extended group structure, a new heavy hermitian and two new non-hermitian gauge bosons must exist 
in the spectrum with masses comparable to the PM fields. In particular, the required hierarchal breaking of the various gauge symmetries as well as the generation of all of the SM and 
PM (Dirac) masses and the mixings necessary to all the PM fermions to decay (unless exotically charged fields are present that are the lightest ones amongst the set of PM) can be accomplished 
via the vevs of just three Higgs (anti-)triplets of $3_L$ having differing $1_A1_B$ quantum numbers. Second, these PM fermion fields necessarily consist of both color singlet as well as color triplet 
states, at least some of which will share $3_L$ (anti-)triplet representations with the SM fermions and so are connected to them via the two new heavy NHGB. PM decays via these NHGB can compete 
with the familiar and usually dominant SM-PM mixing-induced mode provided they are kinematically allowed. Third, at the high scale, the $U(1)_D$ gauge coupling, $g_D$, is found to be bounded 
from above in a manner dependent upon the value of $|\sigma|$ but is always constrained to be $\lsim 0.83 g_L$. This bound can have important implications for low energy dark sector searches 
if $g_D$ runs to even smaller values as the $\sim 1$ GeV scale is approached from above as we might anticipate. Lastly, if kinematically allowed, PM and/or NHGB can appearing as final states in 
resonant $Z'_M$ decay can have enhanced production rates and correspondingly extended mass reaches beyond the typical naive estimates.

PM models with extended gauge sectors linking the dark and SM sectors can lead to a wide spectrum of interesting and complex phenomenology at existing and future colliders. Hopefully signals of the 
physics of the dark sector will soon be observed.

\section*{Acknowledgements}
The author would like to particularly thank J.L. Hewett for valuable discussions and the Brookhaven National Laboratory Theory Group for its great hospitality during several visits. This work was 
supported by the Department of Energy, Contract DE-AC02-76SF00515.



\end{document}